\documentclass[11pt]{article}
\usepackage{amsfonts}
\usepackage[margin=2cm]{geometry}
\usepackage{amsmath,amssymb,amsthm,array,booktabs,cite,dsfont,graphicx,mathtools,multirow,placeins,rotating,stmaryrd,tensor,verbatim}
\usepackage[bookmarksnumbered,linktocpage,pdfstartview=FitH]{hyperref}
\usepackage{hypcap}
\usepackage[margin=2cm]{geometry}
\usepackage{caption}
\usepackage{color}
\usepackage{setspace}
\usepackage{slashed}
\usepackage{changepage}

\newcolumntype{M}[1]{>{$}{#1}<{$}}

\newcommand{\sst}[1]{{\scriptscriptstyle #1}}

\newcommand{\rep}[1]{\ensuremath{\mathbf{#1}}}

\def\0{{\sst{(0)}}}
\def\1{{\sst{(1)}}}
\def\2{{\sst{(2)}}}
\def\3{{\sst{(3)}}}
\def\4{{\sst{(4)}}}
\def\5{{\sst{(5)}}}
\def\6{{\sst{(6)}}}
\def\7{{\sst{(7)}}}

\newcommand{\be}{\begin{equation}}
\newcommand{\ee}{\end{equation}}
\def\ba{\begin{array}}
\def\ea{\end{array}}

\newcommand{\bea}{\begin{eqnarray}}
\newcommand{\eea}{\end{eqnarray}}

\newcommand{\y}[1]{\ensuremath{\tilde{#1}}}

\DeclareMathOperator{\Aut}{Aut}

\DeclareMathOperator{\Hom}{Hom}

\DeclareMathOperator{\SO}{SO}

\DeclareMathOperator{\USp}{USp}
\DeclareMathOperator{\SL}{SL}
\DeclareMathOperator{\SU}{SU}
\DeclareMathOperator{\Sp}{Sp}

\DeclareMathOperator{\Un}{U}

\DeclareMathOperator{\Spin}{Spin}

\newcommand{\N}{\mathcal{N}}

\newcommand{\blf}[2]{\langle#1 , #2\rangle}
\newcommand{\bn}{\mathbf{n}}
\newcommand{\J}{\mathfrak{J}}

\newcommand{\alg}{\mathds{A}}
\newcommand{\mf}{\mathfrak}

\newcommand{\R}{\mathds{R}}
\newcommand{\C}{\mathds{C}}
\newcommand{\Q}{\mathds{H}}
\newcommand{\Oct}{\mathds{O}}
\newcommand{\Z}{\mathds{Z}}
\newcommand{\sigmabar}{\bar{\sigma}}

\newcommand{\tN}{\tilde{\mathcal{N}}}

\newcommand{\tA}{\tilde{A}}
\newcommand{\tG}{\tilde{G}}

\newcommand{\td}[1]{\tilde{#1}}

\begin{document}

\begin{titlepage}
\begin{center}
\hfill DIAS-STP-15-07\\
\hfill IMPERIAL-TP-2015-MJD-01\\
\vskip 2cm

{\huge \bf Gravity as the square of Yang-Mills?\footnote{Lectures delivered by M. J. Duff}}

\vskip 1.5cm

{\bf L.~Borsten${}^{1}$ and M.~J.~Duff${}^2$}

\vskip 20pt
 {\it ${}^1$School of Theoretical Physics, Dublin Institute for Advanced Studies,\\
10 Burlington Road, Dublin 4, Ireland}\\\vskip 5pt
{\it ${}^2$Theoretical Physics, Blackett Laboratory, Imperial College London,\\
 London SW7 2AZ, United Kingdom}\\\vskip 5pt

\texttt{leron@stp.dias.ie}\\
\texttt{m.duff@imperial.ac.uk}\\

\end{center}

\vskip 2.2cm

\begin{center} {\bf ABSTRACT}\\[3ex]\end{center}

In these lectures we review  how the symmetries of gravitational theories  may be regarded as originating from those of ``Yang-Mills squared''. We begin by motivating the idea that certain aspects of gravitational theories can be captured by the product, in some  sense,  of two distinct Yang-Mills theories, particularly in the context of scattering amplitudes. We then introduce a concrete dictionary for  the covariant fields of (super)gravity  in terms of the product of two (super) Yang-Mills theories. The dictionary implies   that the  symmetries of each (super) Yang-Mills factor generate the symmetries of the corresponding (super)gravity theory: general covariance, $p$-form gauge invariance, local Lorentz invariance, 
 local supersymmetry, R-symmetry and U-duality.
\vfill

\end{titlepage}

\newpage \setcounter{page}{1} \numberwithin{equation}{section} \tableofcontents 

\newpage
\section{MJD: Tribute to Dick Arnowitt}

It was thanks to Dick Arnowitt that I spent eleven wonderful years 1988-1999 here at Texas A\&M. My wife Lesley and I have not forgotten the kindness shown by Dick and Young-In when we first arrived here from England.  It was a privilege to work in the first rate Theoretical Physics Group that Dick had built.

\section{Introduction}
The idea that gravitational physics can be understood in terms of gauge theory has reoccured a number of times, in a variety of guises. 
The  most conceptually  straight-forward approach is to regard  gravity as the  gauge  theory of Lorentz, Poincar\'e or de Sitter symmetries    \cite{1956PhRv..101.1597U, 1961JMP.....2..212K,1977PhRvL..38..739M, 1977NuPhB.129...39C, 1979JPhA...12L.205S}. The holographic principle \cite{hooft1994dimensional, Susskind:1994vu},  concretely realised through the AdS/CFT correspondence \cite{Maldacena:1997re,Witten:1998qj,Gubser:1998bc}, represents  a more subtle  realisation of this notion, with profound consequences for our understanding of both gauge and gravity theories. Here we appeal to a third and, at least superficially, independent incarnation:
\be\label{ymsquared}
\text{gravity}=\text{Yang-Mills}\times\text{Yang-Mills}.
\ee
At first sight this is a radical proposal; Einstein's general relativity describes gravity as the dynamics of spacetime, while Yang-Mills theories, as used to describe  the strong, weak and electromagnetic forces, play out \emph{on} spacetime. General relativity  and Yang-Mills theory are seemingly worlds apart in almost every regard, from their fundamental degrees freedom to their basic  symmetries. In particular,  the Yang-Mills theories underlying the standard model are renormalisable, predictive quantum field theories, in stark contrast to perturbative quantum gravity.  

Despite their differences, however, there already exist some fascinating hints that gravity, at least in some regimes, may be related  to the square of Yang-Mills theory. String theory provided the first example  in the form of the Kawai-Lewellen-Tye (KLT) relations, which connect tree-level amplitudes of closed strings to sums of products of open string amplitudes \cite{Kawai:1985xq}.  More recently,  invoking Bern-Carrasco-Johansson  (BCJ) colour-kinematic duality \cite{Bern:2008qj} it has been  conjectured  \cite{Bern:2010ue} that the on-mass-shell momentum-space scattering amplitudes for gravity are the  ``double-copy''  of gluon scattering amplitudes in Yang-Mills theory to all orders in perturbation theory.

The recent  renaissance in amplitude calculations has been principally driven by the ``on-shell paradigm''. Starting with Lagrangian field theory we learnt how to compute simple amplitudes to low  orders in perturbation theory. The factorial growth in complexity with loop order quickly renders traditional approaches impractical. Searching for  computational efficiency, over time  various generic amplitude structures (on-shell recursion relations, generalised unitarity cuts, Grassmannians, scattering equations \ldots) were uncovered,  eventually allowing the Lagrangian ladder to be kicked away. For  an overview of  these developments see \cite{Elvang:2013cua}. This freedom led to the discovery of new features of  amplitudes, not visible from the original Lagrangian   perspective.  BCJ colour-kinematic duality falls into this class of surprises.  Conversely, having climbed so high we can no longer see  where we can from; the full significance  and implications of BCJ duality remain unclear. Can we climb back down by some other route and understand the origin of these remarkable dualities?  The basic idea reviewed here is to build a dictionary expressing the covariant fields of (super) gravity as the product, in a well-defined sense, of two arbitrary (super) Yang-Mills theories.  

 \subsection{Motivation}\label{motivation}
 
 We begin by sketching  the BCJ colour-Kinematic duality and the double copy procedure \cite{Bern:2008qj, Bern:2010ue, Bern:2010yg}. For a more detailed account of this topic the reader is referred to the reviews \cite{Elvang:2013cua, Carrasco:2015iwa}. This will not only better motivate \eqref{ymsquared}, but also inform our field theory constructions in the subsequent sections.

Let us consider the $n$-point $L$-loop  amplitude of Yang-Mills theory  with an arbitrary gauge group. Converting all four-point contact terms into $s$, $t$ or $u$ channel trivalent pole diagrams by inserting propagators $1=s/s=t/t=u/u$, we have, 
\be\label{amp}
A_{n}^{L}=i^Lg^{n-2+2L}\sum_{i\in \text{trivalent graphs}}\int \prod^{L}_{l=1}\frac{d^Dp_l}{(2\pi)^DS_i} \frac{c_i n_i}{\prod_{a_i}p_{a_i}^{2}}.
\ee
The sum is over all $n$-point $L$-loop graphs $i$ with only trivalent vertices. $c_i$ denotes the kinematic factor of graph $i$, composed of gauge group structure constants. $n_i$ denotes the kinematic factor of graph $i$. It is a polynomial of Lorentz-invariant contractions  of polarisation vectors and momenta. The $p_{a_i}^{2}$ are the propagators for each graph $i$. $S_i$ is the dimension of the automorphism group of graph $i$.

The set of $n$-point trivalent graphs can be organised into triples $i, j, k$ such that they differ in only one propagator. For such a triple the three disctint propagators are embedded in the same graph, connected to the same four incoming edges, but in the $s, t, u$ channel  for (say) $i, j, k$, respectively. For such a triple the colour factors will obey a Jacobi identity
\be
c_i+c_j+c_k=0
\ee 
and consequently the generalised gauge transformations 
\be
n_i\rightarrow n_i+s \Delta,\qquad n_j\rightarrow n_j+t \Delta,\qquad n_k\rightarrow n_k+ u \Delta,
\ee
leave the amplitude \eqref{amp} invariant \cite{Bern:2008qj}. It was proposed in \cite{Bern:2008qj} that one can arrange the diagrams, using the generalised gauge transformations   if necessary, to display a colour-kinematic duality:
\be
c_i+c_j+c_k=0\Rightarrow n_i+n_j+n_k=0
\ee
and if $c_i\rightarrow-c_i$ under the interchange of two legs then   $n_i\rightarrow-n_i$.  A reorganisation admitting this surprising  relationship between  colour and kinematic data was shown to exist
  for all $n$-point tree-level amplitudes in \cite{Bern:2010yg}.  Although there is as yet no proof, the colour-kinematic duality is conjectured to hold, with highly non-trivial evidence \cite{Bern:2009kd, Bern:2014sna}, at any loop level, thus going beyond the KLT relations \cite{Bern:2008qj, Bern:2010ue}.  While it is clear that the colour factors should obey Jacobi identities (by definition), it is not at all obvious that the kinematic factors  should play by the same rules! 
  
This suggests that there is in fact some underlying kinematic algebra mirroring   the properties of conventional Lie algebras, as described in \cite{Monteiro:2011pc, Monteiro:2013rya}.     In general, this hidden algebra cannot be made manifest at the Lagrangian level, however for the self-dual sector it can be identified as a diffeomorphism Lie algebra,  which determines the kinematic numerators of generic tree-level maximally helicity violating  amplitudes \cite{Monteiro:2011pc}. Important features of the BCJ construction can also be derived from string theory. In particular, the BCJ relations \cite{Bern:2008qj} (which we have not discussed) have been obtained via monodromy relations \cite{BjerrumBohr:2009rd, Stieberger:2009hq, Tye:2010dd}. Moreover, explicit expressions for colour-kinematic duality respecting local tree-level numerators at $n$-points have been derived using the pure spinor approach to string theory amplitudes and dimensional reduction \cite{Mafra:2011kj, Mafra:2015vca}. The string theoretic approach is also suited to loop level calculations. For example, colour-kinematic duality respecting numerators at five points with one and two  loops were obtained in  \cite{Mafra:2014gja, Mafra:2015mja}, while   at arbitrary multiplicity one-loop maximally helicity violating colour-kinematic duality respecting numerators  have been constructed in \cite{He:2015wgf}.

\begin{figure}\centering
\includegraphics[width=0.75\textwidth]{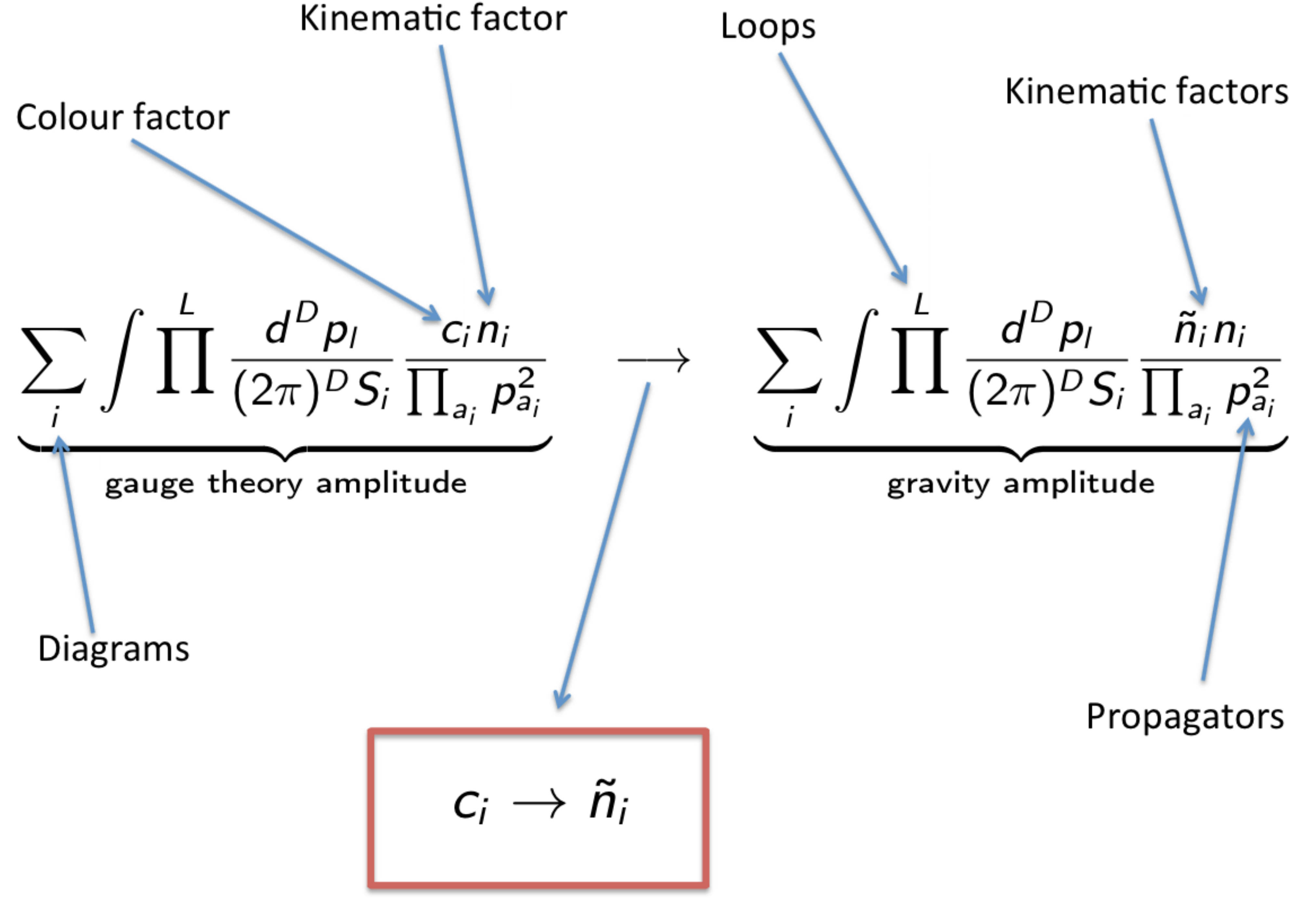}
\caption{The double-copy procedure. Assuming the gauge theory amplitude on the left has been arranged to display colour-kinematic duality then the gravity amplitude on the right is straight-forwardly obtained by replacing the colour factors with a second copy of the kinematic factors. Note, the second factor does not have correspond to the same Yang-Mills theory. The (supressed) Yang-Mills coupling constants must be replaced by the gravitational coupling constant $g\rightarrow \kappa/2$, where $\kappa^2=16\pi G_N$.\label{dc}}
\end{figure}

More remarkable still is the double-copy prescription \cite{Bern:2008qj, Bern:2010yg, Bern:2010ue}. Assuming one has found a colour-kinematic duality respecting representation of the $n$-point $L$-loop gluon amplitude, the equivalent $n$-point $L$-loop graviton amplitude is obtained by simply replacing each colour factor, $c_i$, with a second kinematic factor, $\tilde{n}_i$, as  depicted  in \autoref{dc}. Examining the unitary cuts of the gravity amplitude obtained via the double-copy   is sufficient to prove it reproduces the correct result, assuming colour-kinematic duality is satisfied in one of the Yang-Mills factors. These ideas are seamlessly extended to supersymmetric theories. In particular, the ÒsquareÓ of the amplitudes of the maximally supersymmetric $\N = 4$ super Yang-Mills theory yield amplitudes of the maximally supersymmetric $\N = 8$ supergravity theory.

The double-copy picture is not only  conceptually compelling but also computationally powerful,  bringing  previously intractable calculations with in reach.  This has pushed forward dramatically our understanding of divergences in perturbative  quantum gravity, revealing a number of unexpected features and calling into question previously accepted arguments regarding finiteness. For instance, the four-point graviton amplitude in $\N=8$ supergravity  has been shown to be finite to four loops  \cite{Bern:2009kd}, contradicting some early expectations \cite{Howe:1988qz}. In particular, the existence of a supersymmetric $R^4$ counter-term in $\mathcal{N}=1$ supergravity at three loops was established in \cite{Deser:1977nt}, although it was already noted there that this result may not hold when more supersymmetry is considered. It has since been shown that the four-loop cancellation can be accounted for by supersymmetry and $E_{7(7)}$ U-duality  \cite{Green:2010sp, Bossard:2010bd, Beisert:2010jx, Bossard:2011tq, Bossard:2012xs}. The consensus, however, is that at seven loops any would-be cancellations cannot be ``consequences of supersymmetry in any conventional sense''  \cite{Green:2010sp}. Unfortunately, seven loops in $\N=8$ supergravity remains beyond reach but by decreasing the amount of supersymmetry these arguments apply at lower loop order. Indeed, the four-point amplitude of $D=4, \N=5$ supergravity has been shown to be finite to four loops, contrary to all expectations based on standard symmetry arguments \cite{Bern:2014sna}.  There are ``enhanced cancellations''  at work and the conclusion that $\N=8$ supergravity will diverge at seven loops is thrown into doubt. Although  the majority opinion is that $\N=8$ supergravity will diverge 
at some loop order,  there is something deeper at work we have yet to understand fully and question remains very much open.

\section{Covariant field dictionary}
     These  developments raise the question: to what extent, or in what sense, can one regard gravity as the square of Yang-Mills? Is there a deeper connection underlying the amplitude relations? One approach to addressing such questions is to build a  dictionary at the level of fields, as opposed to on-shell states or amplitudes. In a sense this runs contrary to the ``on-shell paradigm'' that took us here. Going back off-shell may nonetheless  be instructive. This approach has been examined  at the level of Lagrangians in \cite{Bern:2010ue, Monteiro:2011pc} and classical solutions in \cite{Monteiro:2014cda, Luna:2015paa, Luna:2016due}. 
Here instead we focus on expressing the  covariant fields of (super)gravity in terms   of the product of (super) Yang-Mills fields. The first consistency check such a dictionary must pass is at the level of  symmetries. As we shall review, the  gravitational symmetries of general covariance, $p$-form gauge invariance, local supersymmetry and local chiral  symmetry, R-symmetry and U-duality follow from those of Yang-Mills at linearised  approximation.

Much of the squaring literature invokes a mysterious product:
\be
A_\mu(x)``\otimes" \tilde{A}_\nu(x).
\ee
Here, $A_\mu$ and $\tilde{A}_\nu$ are the gauge potentials of two distinct Yang-Mills theories, which we will refer to as left (no tilde) and right (tilde), respectively. They can have arbitrary and independent  non-Abelian gauge groups $G$ and $\tG$. 
 Reading off the meaning of $``\otimes"$ from the tensor product branching rules of the appropriate spacetime little group representations or corresponding string states  one can consistently match the symmetries. See in particular \cite{Siegel:1988qu, Siegel:1995px}. Here, we are instead seeking a concrete definition of  $``\otimes"$ at the level of field theory which is valid whether or not there is an underlying string interpretation. This raises two immediate questions: (i) where do the gauge indices go? (ii) does it obey the Leibnitz rule?

Guided by the structure of the amplitude relations and requirements of symmetry we introduced a covariant product rule in \cite{Anastasiou:2014qba}:
\be\label{def}
f``\otimes" {g}:= f \star \Phi \star {g}.
\ee
Let us review the ingredients in \eqref{def}. The $\star$  product denotes a convolutive inner tensor product with respect to the Poincar\'e group combined with a Killing form $\langle~,~\rangle:\mathfrak{g}\otimes{\mathfrak{g}} \rightarrow \R$,
\be
[f\star g](x)=\int d^Dy\langle f(y), g(x-y)\rangle.
\ee
We have further introduced the ``spectator'' field $\Phi$, a  $G \times \tG$ bi-adjoint valued scalar.    The convolution reflects the fact that the amplitude relations are multiplicative in momentum space. It turns out to be essential for reproducing the local symmetries of (super)gravity from those of the two (super) Yang-Mills factors. The Killing form accounts for the gauge groups, while the spectator field allows for arbitrary and independent   $G$ and $\tG$. It fact, the appearance of $\Phi$ is quite natural from the perspective of amplitude relations. Its necessity was identified by Hodges in the context of twistor-theory \cite{Hodges:2011wm}. From the perspective presented in \autoref{motivation}, rather than sending $c_i\rightarrow \tilde{n}_i$, doubling the kinematics and removing the colour, one could also send $n_i\rightarrow \tilde{c}_i$, doubling the colour and removing the kinematics. In \cite{Cachazo:2013iea, Dolan:2013isa} this was shown at tree-level to yield the amplitudes of a  global $G\times \tG$ bi-adjoint scalar field theory with cubic interaction term,
\be
\mathcal{L}_{\text{int}}=-f_{ijk}\tilde{f}_{i'j'k'}\Phi^{ii'}\Phi^{jj'}\Phi^{kk'}.
\ee
The transformation rules of $\Phi$ are fixed by this  theory. A scalar field also appeared independently, but in close analogy to our spectator field,  in the double-copy construction of Kerr-Schild gravity solutions from Yang-Mills solutions in \cite{Monteiro:2014cda}.

\section{$\N=1$ supergravity}

Having introduced the covariant product, let us now work through the simplest example exhibiting all the local symmetries of interest. We consider the product of a left $\N=1$  and a right $\tilde{\N}=0$  theories at linearized level:
\begin{itemize}
\item  Off-shell ${\cal N}=1$  Yang-Mills multiplet with (4+4) bosonic $+$ fermionic degrees of freedom and gauge group $G$:
 \be
 A_{\mu}, \qquad \psi, \qquad D
 \ee
\item Off-shell  off-shell $\tN=0$ Yang-Mills multiplet with (3+0) bosonic $+$ fermionic degrees of freedom  and  gauge group $\tG$: 
\be
\tilde{A}_{\nu}
\ee
\end{itemize}

Without making any assumptions regarding the dynamics this yields the $(12+12)$ new-minimal ${\cal N}=1$ supergravity multiplet \cite{Sohnius:1981tp}:
\be
g_{\mu\nu}, \qquad B_{\mu\nu}, \qquad  \psi_{\mu}, \qquad  V_{\mu} 
\ee
where general covariance, 2-form gauge invariance, local supersymmetry and local chiral  symmetry follows from the left/right gauge symmetries.

The ``gravity$=$Yang-Mills$\times$Yang-Mills'' dictionary and symmetry transformations are most concisely expressed in the superfield formalism. Hence, we consider:
\begin{enumerate}
\item A left  $\N=1$ real vector superfield,
 \be
\begin{split}
V(x, \theta, \bar{\theta})=&C+i\theta\chi-i{\bar \theta}{\bar \chi}+i{\theta}^2{ F}-i{\bar \theta}^2{\bar F}-{ \theta}\sigma^{\mu}{\bar \theta}A_\mu\\
&+i{ \theta}^2{\bar\theta}\left({\bar{\psi}}+\frac{i}{2}\bar{\sigma}^\rho \partial_\rho{ \chi} \right )
-i{\bar\theta}^2{ \theta}\left(\psi+\frac{i}{2}\sigma^\rho \partial_\rho{\bar \chi} \right)\\
&+\frac{1}{2}{\bar \theta}^2{\theta}^2\left (D +\frac{1}{2} \Box C \right)
\end{split}
\ee
 transforming under local supergauge, non-Abelian global $G$ and  global super-Poincar\'{e}:
 \be
\delta V=\underbrace{\Lambda +\bar \Lambda}_{\text{local Abelian supergauge}}+\overbrace{[V, X]}^{\text{global non-Abelian} ~G}+\underbrace{\delta_{(a, \lambda, \epsilon)} V}_{\text{global super-Poincar\'{e}}}
\ee
 where $\Lambda(x, \theta, \bar{\theta})$ is a chiral superfield of supergauge parameters
\be
\Lambda(x, \theta, \bar{\theta}) = B+\sqrt{2}\theta \zeta + \theta^2 K +i{ \theta}\sigma^{\rho}{\bar \theta}\partial_{\rho}a
 +\frac{i}{\sqrt{2}}\theta^2{\bar\theta}{\bar\sigma}^\rho\partial_\rho\zeta+\frac{1}{4}\theta^2{\bar\theta}^2\Box B
\ee 
\item A right $\tN=0$ Yang-Mills potential  
$
\tA_{\nu}
$ 
transforming under local gauge, non-Abelian global $\tG$ and  global Poincar\'{e}:
\[
\delta \tA_\nu=\underbrace{\partial_\nu\td{\sigma}}_{\text{local Abelian gauge}}+\overbrace{[\tA_\nu, \td{X}]}^{\text{global non-Abelian} ~\tG}+\underbrace{\delta_{(a, \lambda)}A_\nu}_{\text{global Poincar\'{e}}}
\]
\item The spectator bi-adjoint scalar $\Phi$ field transforming under non-Abelian global $G\times \tG$  and global Poincar\'{e}:
\be
\delta\Phi=\overbrace{-[\Phi, X]-[\Phi, \td{X}]}^{\text{global non-Abelian} ~G\times \tG}+\underbrace{\delta_{a}\Phi}_{\text{global Poincar\'{e}}}
\ee
\end{enumerate}
The gravitational  symmetries are reproduced here from those of Yang-Mills by invoking the gravity/Yang-Mills dictionary for fields and supergauge parameters:
\be
\begin{array}{l|lllllllllllll}
\text{Fields}  & \varphi_{\nu}& =& V &\star& \Phi& \star& \tA_\nu & \text{real superfield} \\[5pt]
\hline
\\
\text{Paras}  & \phi&=&V& \star &\Phi& \star& \y{\lambda}&  \text{real superfield} \\  [5pt]
&S_\nu&=& \Lambda &\star &\Phi &\star &\tA_\nu& \text{chiral superfield} \\[5pt]
 \end{array}
 \ee

Varying the gravitational superfield  
\be\label{nm}
\begin{split}
\varphi_{\nu}(x, \theta, \bar{\theta})=&C_\nu+i\theta\chi_\nu-i{\bar \theta}{\bar \chi}_\nu+i{\theta}^2{ F}_\nu-i{\bar \theta}^2{\bar F}_\nu-{ \theta}\sigma^{\mu}{\bar \theta}(g_{\mu\nu}+B_{\mu\nu})\\
&+i{ \theta}^2{\bar\theta}\left(\bar{\psi}_\nu+\frac{i}{2}\bar{\sigma}^\rho \partial_\rho{ \chi}_\nu \right )
-i{\bar\theta}^2{ \theta}\left( \psi_\nu+\frac{i}{2}\sigma^\rho \partial_\rho{\bar \chi}_\nu \right)\\
&+\frac{1}{2}{\bar \theta}^2{\theta}^2\left ( V_\nu +\frac{1}{2} \Box C_{\nu} \right)
\end{split}
\ee
via the dictionary
\be
\delta \varphi_{\nu}= \delta V\star \Phi\star \tA_\nu
+V\star \delta \Phi\star \tA_\nu
+V\star \Phi\star \delta \tA_\nu
\ee
 we obtain
\be\label{trans}
\delta \varphi_{\nu} ={S_\nu+{\bar S}_\nu+{\partial}_{\nu}\phi}+\delta_{(a, \lambda, \epsilon)} \varphi_{\nu}.
\ee
This is the complete set of transformation rules for the new-minimal superfield at linearised approximation. Note, this derivation makes use of 
\be
\langle [X, Y], Z\rangle= \langle X, [Y, Z]\rangle 
\ee
and, crucially, the convolution property
\be
 \partial_\mu (f\star g)=(\partial_{\mu} f)\star g=f \star (\partial_{\mu}g).
\ee
To summarise, we have obtained  the field content \eqref{nm} and transformation rules \eqref{trans} at linearised approximation of new-minimal   $\N=1$ supergravity \cite{Cecotti:1987qe, Ferrara:1988qxa}. Hence, 
 the local gravitational symmetries of general covariance, 2-form gauge invariance, local supersymmetry and local chiral  symmetry follow from those of Yang-Mills at linear level.
 
 Introducing field equations we should match the on-shell content of the tensor product of spacetime little group representations. This is done covariantly  by including the ghost sector in the dictionary \cite{Siegel:1988qu,Siegel:1995px}. The  $12+12$ multiplet   splits  with respect to superconformal transformations into an $8+8$ conformal supergravity multiplet plus a $4+4$ conformal tensor multiplet,
\be
\underbrace{\begin{pmatrix}
\mathbf{5+3+1+3}\\
\mathbf{4+2+4+2}\\
\end{pmatrix}}_{\text{new-minimal}}
\rightarrow \underbrace{\begin{pmatrix}
\mathbf{5+3}\\
\mathbf{4+4}\\
\end{pmatrix}}_{\text{conformal}}+\underbrace{\begin{pmatrix}
\mathbf{3+1}\\
\mathbf{2+2}\\
\end{pmatrix}}_{\text{tensor}}
\ee
in terms of  $\SO(3)$ representions. 
Since the left (anti)ghost is a chiral superfield the ghost-antighost sector  gives a compensating $4+4$ chiral (dilaton) multiplet  \cite{Siegel:1988qu,Siegel:1995px}, yielding   old-minimal  $12+12$ supergravity \cite{Stelle:1978ye, Ferrara:1978em}  coupled to a tensor multiplet, which, with the conventional 2-derivative Lagrangian,   correctly corresponds to the on-shell content obtained by tensoring left/right helicity states.

\section{Extended supersymmetry and U-duality}

This minimally supersymmetric  example does not fully address the issue of U-duality \cite{Hull:1994ys}, which, in context of string/M-theory, is of fundamental importance.   U-duality manifests itself in supergravity, the low energy effective limit of string/M-theory, in the form of non-compact global symmetries, $\mathcal{G}$, acting non-linearly on the scalar fields \cite{Cremmer:1979up}. In all cases obtained from  ``Yang-Mills$\times$Yang-Mills'' the scalars parametrise a symmetric space $\mathcal{G/H}$, where $\mathcal{H}$ is the maximal compact subgroup of $\mathcal{G}$ \cite{Anastasiou:2015vba}. The U-dualities and corresponding global symmetries for M-theory compactified on a $n$-torus are summarised in \autoref{uduality}. Note, we will also use the term U-duality to refer to $\mathcal{G}$. The question of global symmetries from squaring Yang-Mills has also been addressed in \cite{Bianchi:2008pu, Chiodaroli:2011pp, Carrasco:2012ca, Chiodaroli:2014xia}, particularly in the context of scattering amplitudes.

\begin{table}[ht]
\begin{tabular*}{\textwidth}{@{\extracolsep{\fill}}cM{c}M{c}M{c}c}
\hline
\hline
 $n$-torus   & \text{U-duality}  & \mathcal{G}                                         & \mathcal{H}                                        & \\
\midrule
 1   & \SO(1,1,\mathds{Z})              & \SO(1,1,\mathds{R})                        & -                                        & \\
 2     & \SL(2,\mathds{Z})\times \SO(1,1,\mathds{Z})          & \SL(2,\mathds{R})\times \SO(1,1,\mathds{R}) & \SO(2,\mathds{R})                         & \\
 3     & \SL(2,\mathds{Z})\times \SL(3,\mathds{Z})           & \SL(2,\mathds{R})\times \SL(3,\mathds{R})   & \SO(2,\mathds{R})\times \SO(3,\mathds{R})  & \\
 4     & \SL(5,\mathds{Z})          & \SL(5,\mathds{R})                          & \SO(5,\mathds{R})                         & \\
 5     & \SO(5,5,\mathds{Z})           & \SO(5,5,\mathds{R})                        & \SO(5,\mathds{R})\times SO(5,\mathds{R})  & \\
 6     & E_{6(6)}(\mathds{Z})         & E_{6(6)}(\mathds{R})                      & \USp(8)                                   & \\
 7     & E_{7(7)}(\mathds{Z})             & E_{7(7)}(\mathds{R})                      & \SU(8)                                    & \\
 8     & E_{8(8)}(\mathds{Z})         & E_{8(8)}(\mathds{R})                      & \SO(16,\mathds{R})                        & \\
\hline
\hline
\end{tabular*}
\caption{U-dualities  (global symmetries) of M-theory ($D=11, \N=1$ supergravity) compactified on an $n$-torus.}\label{uduality}
\end{table}

As made clear by \autoref{uduality}, U-duality becomes increasingly manifest as one descends in dimension\footnote{We stop at $D=3$, which has $E_{8(8)}$ U-duality, the largest finite dimensional exceptional Lie algebra. One can continue to $D=2,1,0$, invoking the infinite dimensional extended algebras $E_{9(9)}, E_{10(10)}, E_{11(11)}$ \cite{Julia:1980gr, Nicolai:1987kz, West:2001as, Damour:2002cu}. Although we will not discuss theses cases here, it would be interesting to investigate whether they can be understood from the perspective of Yang-Mills squared.}. Thus, to fully  expose the structure of U-duality with respect to squaring  we should consider the product in  $D=3$ of left  Yang-Mills theories with $\N=1,2,4,8$  and right Yang-Mills theories with $\tilde{\N}=1,2,4,8$. This was done in \cite{Borsten:2013bp}. The result revealed a rather intriguing mathematical structure. The U-duality algebras obtained make up the Freudenthal-Rosenfeld-Tits magic square \cite{Freudenthal:1954, Tits:1955, Rosenfeld:1956} as given in \autoref{ymsquare}. As we shall explain this surprise has an elegant explanation, but first we must spend some time on the magic square itself.

Note, the real forms appearing in \autoref{ymsquare} are not unique;  there are numerous possibilities as described in \cite{Cacciatori:2012cb}. They also play a role in supergravity. In particular, the $\C, \Q,$ and $\Oct$ rows of one such magic square (distinct from \autoref{ymsquare}) describe the U-dualities of the aptly named magic supergravities in $D = 5, 4, 3$ respectively \cite{Gunaydin:1983rk, Gunaydin:1983bi, Gunaydin:1984ak}. It should be emphasised, however, that the appearance  of the magic square here is unrelated to these constructions.

 \begin{table}[ht]
 \begin{center}
\begin{tabular}{c|cccccc}
\hline
\hline
$\N \otimes \tilde{N}$ && 1 & 2  & 4  & 8  \\
 \hline
 \\
  1 && $\mathfrak{sl}(2, \R)$ & $\mathfrak{su}(2,1)$   & $\mathfrak{sp}(4,2)$   & $\mathfrak{f}_{4(-20)}$   \\
  2& &$\mathfrak{su}(2,1)$ & $\mathfrak{su}(2,1)\times  \mathfrak{su}(2,1)$   & $\mathfrak{su}(4,2)$   & $\mathfrak{e}_{6(-14)}$   \\
  4 & &$\mathfrak{sp}(4,2)$ & $\mathfrak{su}(4,2)$   & $\mathfrak{so}(8,4)$   & $\mathfrak{e}_{7(-5)}$   \\
   8 & &$\mathfrak{f}_{4(-20)}$ & $\mathfrak{e}_{6(-14)}$   & $\mathfrak{e}_{7(-5)}$   & $\mathfrak{e}_{8(8)}$   \\
   \\
   \hline
   \hline
\end{tabular}
\caption{The magic square of U-duality algebras obtained from the product of two Yang-Mills theories in $D=3$ spacetime dimensions.}\label{ymsquare}
\end{center}
\end{table}

\subsection{Division algebras and the magic square}
In this section we follow closely \cite{Barton:2003, Baez:2001dm}; we refer the reader to these works for more detailed explanations and proofs. An algebra $\mathds{A}$ defined over  $\R$ with identity element $e_0$, is said to be \emph{composition} if it has a non-degenerate quadratic form\footnote{A \emph{quadratic norm} on a vector space $V$ over a field $\R$ is a map $\bn:V\to\R$ such that: (1) $\bn(\lambda a)=\lambda^2\bn(a),  \lambda\in\R, a\in V$ and 
(2)
$
\langle a, b\rangle:=\bn(a+b)-\bn(a)-\bn(b)
$
is bilinear.}
$\bn:\mathds{A}\to\R$ such that,
\begin{equation}
\bn(ab)=\bn(a)\bn(b),\quad \forall~~ a,b \in\alg,
\end{equation}
where we denote the multiplicative product of the algebra by juxtaposition.  Regarding ${\R}\subset\alg$ as the scalar multiples of the identity $ \R e_0$ we may decompose $\mathds{A}$ into its ``real'' and ``imaginary'' parts $\alg={\R}\oplus \alg'$, where $\alg'\subset\alg$ is the subspace orthogonal to $\R$. An arbitrary element $a\in\alg$ may be written $a=\text{Re}(a) +\text{Im}(a)$. Here  $\text{Re}(a)\in\R e_0$, $\text{Im}(a)\in \alg'$ and
\be
\text{Re}(a)=\frac{1}{2}(a+\overline{a}), \qquad \text{Im}(a)=\frac{1}{2}(a-\overline{a}),
\ee
where we have defined conjugation  using the bilinear form,
\be
\overline{a}:=\blf{a}{e_0}e_0-a, \qquad \langle a, b\rangle:=\bn(a+b)-\bn(a)-\bn(b).
\ee 

A composition algebra $\mathds{A}$  is said to be \emph{division} if it contains no zero divisors,
\begin{equation*}
ab=0\quad \Rightarrow\quad a=0\quad\text{or}\quad b=0,
\end{equation*}
in which case $\bn$ is positive semi-definite and $\alg$ is referred to as a normed division algebra. Hurwitz's celebrated theorem states that there are exactly four normed division algebras \cite{Hurwitz:1898}: the reals, complexes, quaternions and octonions, denoted respectively by $\R, \C, \Q$ and $\Oct$. They may be constructed via the Cayley-Dickson doubling procedure, $\alg'=\alg\oplus\alg$  with multiplication in $\alg'$ defined by
\be
(a, b)(c, d) = (ac - d\bar{b}, \bar{a}d + cb).
\ee
With each doubling a property is lost as summarised here:
\[
\begin{array}{lllllll}
\alg & Construction & Dim& Division & Associative & Commutative & Ordered \\
\R & \R & 1& yes & yes & yes & yes \\
\C &\R \oplus \R & 2& yes & yes & yes & no \\
\Q &\C \oplus  \C & 4& yes & yes & no & no \\
\Oct &\Q \oplus  \Q & 8& yes & no & no & no \\
\mathds{S} & \Oct\oplus  \Oct & 16 & no & no & no & no \\
\end{array}
\]
On doubling the octonions, $\mathds{S} \cong\Oct\oplus  \Oct$,  the division property fails and we will not consider such cases here. Note that, while the octonions are not associative they are alternative:
\be
[a, b, c]:= (ab)c-a(bc)
\ee
is an alternating function under the interchange of its arguments. This property is crucial for supersymmetry.

\begin{figure}[h!]
  \centering
    \includegraphics[width=0.4\textwidth]{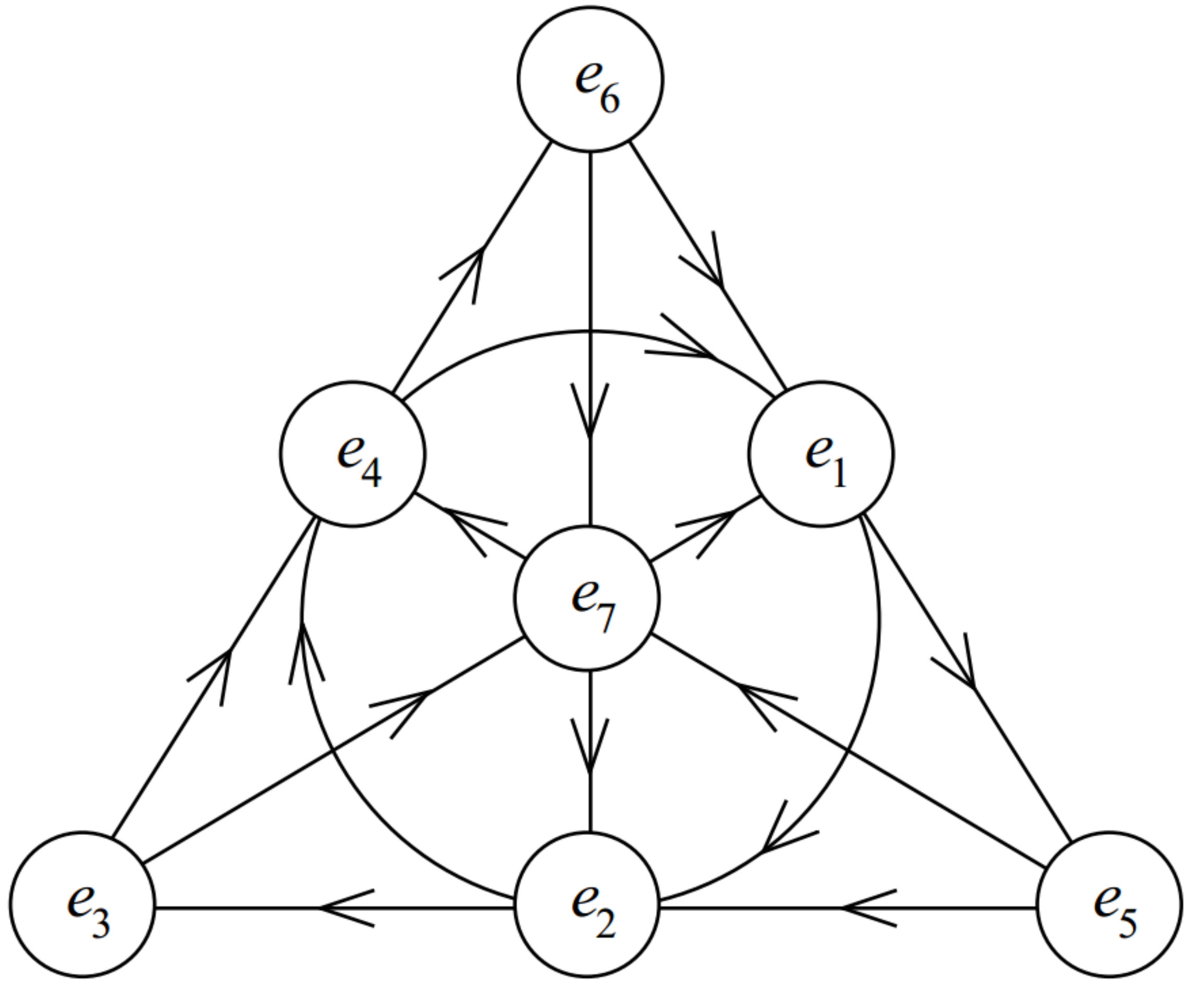}
  \caption{\footnotesize{The Fano plane. The structure constants are determined by the Fano plane, $C_{ijk}=1$ if $ijk$ lies on a line and is ordered according as its orientation. Each oriented line follows the rules of quaternionic multiplication. For example, $e_2e_3=e_5$ and cyclic permutations; odd permutations go against the direction of the arrows on the Fano plane and we pick up a minus sign, e.g. $e_3e_2=-e_5$.}}\label{FANO}
\end{figure}

An element $a\in\Oct$ may be written $a=a^ae_a$, where $a=0,\ldots,7$,  $a^a\in \R$ and $\{e_a\}$ is a basis with one real $e_0$ and  seven $e_i, i=1,\ldots, 7,$ imaginary elements. The octonionic multiplication rule is,
\be
e_ae_b=\left(\delta_{a0}\delta_{bc}+\delta_{0b}\delta_{ac}-\delta_{ab}\delta_{0c}+C_{abc}\right)e_c,
\ee
where $C_{abc}$ is totally antisymmetric and $C_{0bc}=0$.
The non-zero $C_{ijk}$  are given by the Fano plane. See \autoref{FANO}.

There are three symmetry algebras on $\alg$ that we will make use of:
\begin{enumerate}
\item The \emph{norm preserving} algebra is defined as,
\be
\mathfrak{so}(\alg):=\{A\in\Hom_{\R}(\alg) | \langle A a, b\rangle+\langle a, A b\rangle=0, \;\forall a, b \in \alg\},
\ee
yielding,
\be
\begin{split}
\mf{so}(\R)&\cong \emptyset, \\
\mf{so}(\C)&\cong\mf{so}(2),\\
\mf{so}(\Q)&\cong \mf{so}(3)\oplus \mf{so}(3),\\
\mf{so}(\Oct)&\cong  \mf{so}(8).
\end{split}
\ee
\item The \emph{triality} algebra of $\alg$ is defined as,
\be\label{TRIDEF}
\mathfrak{tri}(\alg):=\{(A, B, C) \in  \mathfrak{so}(\alg)\oplus\mathfrak{so}(\alg)\oplus\mathfrak{so}(\alg) | A(ab)=B(a)b+aC(b), \; \forall a,b \in \alg\},\ee
yielding,
\be
\begin{split}
\mf{tri}(\R)&\cong \emptyset, \\
\mf{tri}(\C)&\cong\mf{so}(2)\oplus \mf{so}(2),\\
\mf{tri}(\Q)&\cong \mf{so}(3)\oplus \mf{so}(3)\oplus \mf{so}(3),\\
\mf{tri}(\Oct)&\cong  \mf{so}(8).
\end{split}
\ee  
\item One can regard the triality algebra as a generalised form of the \emph{derivation} algebra defined as,
\be
\mathfrak{der}(\alg)=\{A\in\Hom_\R(\alg) | A(ab) = A(a)b+aA(b)\},
\ee
which for $\alg=\Oct$ gives the smallest exceptional Lie algebra, 
\be
\begin{split}
\mf{der}(\R)&\cong \emptyset, \\
\mf{der}(\C)&\cong\emptyset,\\
\mf{der}(\Q)&\cong \mf{so}(3),\\
\mathfrak{der}(\Oct)&\cong  \mathfrak{g}_{2(-14)}.
\end{split}
\ee  
\end{enumerate}

This provides the first example of a division algebraic description of an exceptional Lie algebra. In fact, the entire Freudenthal magic square can be realised in terms of the division algebras. The magic square was the result of an effort to give a unified and geometrically motivated description of Lie algebras, including the remaining exceptional cases of  $\mathfrak{f}_4, \mathfrak{e}_6, \mathfrak{e}_7, \mathfrak{e}_8$. The classical Lie algebras $\mathfrak{so}(n), \mathfrak{su}(2), \mathfrak{sp}(n)$ are very naturally captured  by $\R, \C, \Q$ geometrical structures, respectively. There are a number of ways of articulating this idea, but perhaps the most concise is in terms of the isometries of projective geometries:
 \be\label{isomclassical}
\mathfrak{Isom}(\R\mathds{P}^n)\cong \mathfrak{so}(n+1), \quad \mathfrak{Isom}(\C\mathds{P}^n)\cong \mathfrak{su}(n+1),\quad
\mathfrak{Isom}(\Q\mathds{P}^n)\cong \mathfrak{sp}(n+1).
\ee
This sequence is rather suggestive;  we might expect octonionic projective geometries to yield  exceptional Lie algebras. Despite non-associativity it was shown by Moufang \cite{Moufang:1933}  that one can consistently construct the octonionic projective line and plane, denoted $\Oct\mathds{P}^1$ and $\Oct\mathds{P}^2$, respectively.   The latter is often referred to as the Cayley plane. We cannot go beyond $n=2$ for the octonions\footnote{One way to understand this is in terms of Jordan algebras. Points in $\Oct\mathds{P}^2$ are bijectively identified with trace 1 projectors in $\mathfrak{J}_{3}^{\Oct}$,  the Jordan algebra of $3\times 3$ octonionic Hermitian matrices. However, for $m>3$, $m\times m$ octonionic Hermitian matrices do not form a Jordan algebra.}, which in this context reflects the fact that  there is indeed just a finite set of exceptional Lie algebras not belonging to any  countably infinite family.  The $\Oct\mathds{P}^1$  example is constructed in direct analogy with the real, complex and quaternionic cases\footnote{Non-associativity, however, implies that the line through the origin containing the point $(a, b)$ is not  given by $\{(\alpha a, \alpha b) | \alpha\in \Oct\}$, unless $x=1$ or $y=1$. This obstacle is easily avoided as all non-zero octonions have an inverse; $(a, b)$ is equivalent to $(b^{-1}a, 1)$ or $(1, a^{-1}b)$ for $b\not=0$ or $a\not=0$, giving two charts with a smooth transition function on their overlap. See \cite{Baez:2001dm}.}. It is diffeomorphic to $S^8$. The octonionic plane has a more intricate structure. An element $(a, b, c)\in\Oct^3$ with $\mathbf{n}(a)+\mathbf{n}(a)+\mathbf{n}(c)=1$ and $(ab)c=a(bc)$ gives a point in $\Oct\mathds{P}^2$ (the line through the origin containing $(a, b, c)$ in $\Oct^3$). It is not difficult to show the space of such elements is a 16-dimensional real manifold embedded in $\Oct^3$ through  eight real constraints: $\mathbf{n}(a)+\mathbf{n}(a)+\mathbf{n}(c)=1$ and $(ab)c=a(bc)$.  The lines in $\Oct\mathds{P}^2$ are  copies of $\Oct\mathds{P}^1$ and there is a duality relation  sending lines/points into points/lines preserving the incidence structure.    Borel showed that $F_{4(-52)}$ is the isometry group of a 16-dimensional projective plane, which is none other than $\Oct\mathds{P}^2$. One can show that the points and lines in $\Oct\mathds{P}^2$ are in one-to-one incidence preserving correspondence with trace 1 and 2 projectors in the Jordan algebra of $3\times 3$ octonionic Hermitian matrices $\mathfrak{J}_{3}({\Oct})$ (treating projectors as propositions the incidence relation in $\mathfrak{J}_{3}^{\Oct}$ is given by implication) \cite{Jordan:1949}. Then $F_{4(-52)}=\text{Isom}(\Oct\mathds{P}^2)$ follows automatically from the result of Chevalley and Schafer that $F_{4(-52)}=\Aut(\mathfrak{J}_{3}({\Oct}))$, the group preserving the Jordan product with Lie algebra $\mathfrak{der}(\mathfrak{J}_{3}({\Oct}))$ \cite{Chevally:1950}.  In summary, the sequence in \eqref{isomclassical} is continued to include,
\be\label{f4}
\mathfrak{Isom}(\Oct\mathds{P}^2)\cong \mathfrak{der}(\mathfrak{J}_{3}({\Oct}))\cong\mathfrak{f}_{4(-52)}.
\ee
Since  $F_{4(-52)}$ acts transitively on the space of trace 1 projectors and the stabiliser of a given trace 1 projector is isomorphic to $\Spin(9)$ we have,
\be
\Oct\mathds{P}^2\cong F_{4(-52)}/\Spin(9).
\ee
The Cayley plane is a homogenous symmetric space with $T_p(\Oct\mathds{P}^2)\cong \Oct^2$, which carries the spinor representation of $\Spin(9)$; under $F_{4(-52)}\supset\Spin(9)$ we have 
\be
\rep{52}\rightarrow \rep{36}+\rep{16},
\ee
or in a more division algebraic form,
\be
\begin{split}
\mathfrak{f}_{4(-52)}&\cong \mathfrak{so}(9)+\rep{16}\\
&\cong \mathfrak{so}(\R\oplus\Oct)+\Oct^2\\
&\cong \mathfrak{so}(\Oct)+\Oct+\Oct+\Oct.
\end{split}
\ee
The three $\Oct$ terms in the final line transform in the three triality related 8-dimensional representations of $\mathfrak{so}(8)$, the vector, spinor and conjugate spinor. It is this triality relation which implies that $\mathfrak{tri}(\Oct)\cong\mf{so}(\Oct)$.

Seemingly inspired by the trivial identity $\Oct\cong \R\otimes\Oct$ Boris Rosenfeld \cite{Rosenfeld:1956} proposed a natural extension of this construction,
\be
\mathfrak{Isom}((\C\otimes\Oct)\mathds{P}^2)\cong \mathfrak{e}_{6(-78)},\qquad \mathfrak{Isom}((\Q\otimes\Oct)\mathds{P}^2)\cong \mathfrak{e}_{7(-133)},\qquad \mathfrak{Isom}((\Oct\otimes\Oct)\mathds{P}^2)\cong \mathfrak{e}_{8(-248)},
\ee
thus giving a uniform geometric description for all Lie algebras. The would-be tangents spaces $(\alg\otimes\Oct)^2$ have the correct dimensions and representation theoretic properties. However,  it is not actually possible to construct projective spaces over $\Q\otimes\Oct$ and $\Oct\otimes\Oct$ using the logic applied to $\Oct\mathds{P}^2$, essentially because they  do not yield Jordan algebras, unlike $\C\otimes\Oct$. They nonetheless  can be identified  with Riemannian geometries with isometries $E_{7(-133)}$ and $E_{8(-248)}$,  respectively.  Indeed, the Lie algebra decompositions\footnote{Note, the additional factors are given by  intermediate algebras: $\mathfrak{tri}(\alg)/\mathfrak{int}(\alg)=\emptyset, \mathfrak{u}(1), \mathfrak{sp}(1), \emptyset$ for $\alg=\R, \C, \Q, \Oct$ \cite{Barton:2003}.},
\be
\begin{split}
\mathfrak{f}_{4(-52)}&\cong \mathfrak{so}(\R\oplus\Oct)+(\R\otimes \Oct)^2\\
\mathfrak{e}_{6(-78)}&\cong \mathfrak{so}(\C\oplus\Oct)\oplus\mathfrak{u}(1)+(\C\otimes \Oct)^2\\
\mathfrak{e}_{7(-133)}&\cong \mathfrak{so}(\Q\oplus\Oct)\oplus\mathfrak{sp}(1)+(\Q\otimes \Oct)^2\\
\mathfrak{e}_{8(-248)}&\cong \mathfrak{so}(\Oct\oplus\Oct)+(\Oct\otimes \Oct)^2
\end{split}
\ee
naturally suggest the identifications 
\be
\begin{split}
\text{Isom}((\R\otimes\Oct)\mathds{P}^2)&= F_{4(-52)}/\Spin(9)\\
\text{Isom}((\C\otimes\Oct)\mathds{P}^2)&= E_{6(-78)}/[(\Spin(10)\times\Un(1))/\Z_4]\\
\text{Isom}((\Q\otimes\Oct)\mathds{P}^2)&= E_{7(-133)}/[(\Spin(10)\times\Sp(1))/\Z_2]\\
\text{Isom}((\Oct\otimes\Oct)\mathds{P}^2)&= E_{8(-248)}/[\Spin(16)/\Z_2]
\end{split}
\ee
with  tangent spaces $(\R\otimes \Oct)^2, (\C\otimes \Oct)^2, (\Q\otimes \Oct)^2, (\Oct\otimes \Oct)^2$  carrying the appropriate spinor representations. Using the Tits' construction \cite{Tits:1955} the isometry algebras are given by the natural generalisation of \eqref{f4},
\be
\begin{split}
\mathfrak{f}_{4(-52)}&\cong \mathfrak{der}(\R)\oplus\mathfrak{der}(\mathfrak{J}_{3}({\Oct}))+\text{Im}\R \otimes \mathfrak{J}'_{3}({\Oct})\\
\mathfrak{e}_{6(-78)}&\cong \mathfrak{der}(\C)\oplus\mathfrak{der}(\mathfrak{J}_{3}({\Oct}))+\text{Im}\C \otimes \mathfrak{J}'_{3}({\Oct})\\
\mathfrak{e}_{7(-133)}&\cong \mathfrak{der}(\Q)\oplus\mathfrak{der}(\mathfrak{J}_{3}({\Oct}))+\text{Im}\Q \otimes \mathfrak{J}'_{3}({\Oct})\\
\mathfrak{e}_{8(-248)}&\cong \mathfrak{der}(\Oct)\oplus\mathfrak{der}(\mathfrak{J}_{3}({\Oct}))+\text{Im}\Oct \otimes \mathfrak{J}'_{3}({\Oct}),
\end{split}
\ee
where $\mathfrak{J}'$ denotes the subset of traceless elements in $\J$. Generalising further, the Tits' construction defines a Lie algebra,  
\be\label{tc}
\mathfrak{M}(\alg_1, \alg_2):=\mathfrak{der}(\alg_1)\oplus\mathfrak{der}(\mathfrak{J}_{3}({\alg_2}))+\text{Im}\alg_1 \otimes \mathfrak{J}'_{3}(\alg_2),
\ee 
for an arbitrary pair $\alg_1, \alg_2=\R, \C, \Q, \Oct$, which yields the (compact) magic square given in \autoref{compms}. The ``magic'' is that \autoref{compms} symmetric about the diagonal despite the apparent asymmetry of \eqref{tc}.
 \begin{table}[ht]
 \begin{center}
\begin{tabular}{c|cccccc}
\hline
\hline
$\otimes$ && $\R$ & $\C$  & $\Q$  & $\Oct$  \\
 \hline
 \\
   $\R$ && $\mathfrak{su}(2)$ & $\mathfrak{su}(3)$   & $\mathfrak{sp}(6)$   & $\mathfrak{f}_{4(-52)}$   \\
  $\C$ & &$\mathfrak{su}(3)$ & $\mathfrak{su}(3)\times  \mathfrak{su}(3)$   & $\mathfrak{su}(6)$   & $\mathfrak{e}_{6(-78)}$   \\
  $\Q$ & &$\mathfrak{sp}(6)$ & $\mathfrak{su}(6)$   & $\mathfrak{so}(12)$   & $\mathfrak{e}_{7(-133)}$   \\
   $\Oct$ & &$\mathfrak{f}_{4(-52)}$ & $\mathfrak{e}_{6(-78)}$   & $\mathfrak{e}_{7(-133)}$   & $\mathfrak{e}_{8(-248)}$   \\
   \\
   \hline
   \hline
\end{tabular}
\caption[Magic square of required real forms.]{The magic square given by the Tits' construction.\label{compms}}
\end{center}
\end{table}
To obtain a magic square with the non-compact real forms that follow from squaring Yang-Mills, as given in \autoref{ymsquare}, one can use a Lorentzian Jordan algebra \cite{Cacciatori:2012cb},
\be\label{tclorentzian}
\mathfrak{M'}(\alg_1, \alg_2):=\mathfrak{der}(\alg_1)\oplus\mathfrak{der}(\mathfrak{J}_{1, 2}({\alg_2}))+\text{Im}\alg_1 \otimes \mathfrak{J}'_{1, 2}(\alg_2).
\ee
Later we shall see that Yang-Mills squared gives an alternative form of \eqref{tclorentzian}, based on the Barton-Sudbery triality construction \cite{Barton:2003}, that is manifestly symmetric in $\alg_1, \alg_2$ \cite{Borsten:2013bp, Anastasiou:2013hba}. This symmetric form reflects the fact that the squaring procedure is itself symmetric on interchanging the left and right theories.

\subsection{Division algebras and Yang-Mills theories}

In the two previous sections we saw that the ``square'' of $D=3$ super Yang-Mills theories and the ``square'' of division algebras both led to the magic square of Freudenthal. Surely this is no coincidence. Indeed, there is a long history of work connecting supersymmetry, spacetime and the division algebras \cite{Gunaydin:1975mp, Gunaydin:1976vq, Gursey:1978et, Gunaydin:1979df, Kugo:1982bn, Gunaydin:1983rk, Gunaydin:1983bi, Gunaydin:1984ak, Sudbery:1984, Sierra:1987, Gursey:1987mv,  Green:1987sp, Evans:1987tm, Duff:1987qa,Blencowe:1988sk, Gunaydin:1992zh, Berkovits:1993hx,  Manogue:1993ja, Evans:1994cn, Schray:1994ur, gursey1996role, Manogue:1998rv, Gunaydin:2000xr, Baez:2001dm,Toppan:2003yx,  Gunaydin:2005zz,  Borsten:2008wd,Baez:2009xt, Baez:2010ye, Huerta:2011ic, Huerta:2011aa,  Cacciatori:2012cb, Anastasiou:2014zfa, Huerta:2014loa, Marrani:2014qia}, which as we shall review underlies this magical meeting.

Perhaps the most direct link from division algebras to  spacetime symmetries comes via the Lie algebra isomorphism of Sudbery \cite{Sudbery:1984},
\be
\mf{sl}(2, \alg)\cong\mf{so}(1, 1+\dim\alg),
\ee
which identifies $D=3,4,6,10$ as algebraically special.
This is itself tied to the earlier observation of Kugo and Townsend \cite{Kugo:1982bn} that the existence of minimal super Yang-Mills mulitplets in only $D=3,4,6,10$ is related to the uniqueness of  $\R, \C, \Q, \Oct$. This was followed-up by a number of authors \cite{Chung:1987in, Fairlie:1987td, Manogue:1989ey, Schray:1994fc, dray2000octonionic}, sharpening the correspondence between supersymmetry and division algebras. The final case of $D=10, \alg=\Oct$ was developed most carefully in \cite{Baez:2009xt}, where the link between supersymmetry and the alternativity of $\Oct$ was emphasised. 

Pulling together these ideas, it was shown in \cite{Anastasiou:2013cya} that $\N$-extended super Yang-Mills theories in $D = n + 2$ dimensions are completely specified  (the field content, Lagrangian and transformation rules)  by selecting an ordered pair of division algebras:  $\alg_n$ for the spacetime dimension and  $\alg_{n\N}$ for the degree of supersymmetry, where the subscripts denote the dimension of the algebras.

Consequently, the dual appearances of the magic square in $D=3$, or equivalently for $\alg_n=\R$, can be explained by the observation that  $D = 3, \N = 1, 2, 4, 8$ Yang-Mills theories can be formulated with a single Lagrangian and a single set of transformation rules, using fields valued in $\R, \C, \Q$ and $\Oct$, respectively \cite{Borsten:2013bp}. Tensoring an  $\alg$-valued $D=3$ super Yang-Mills multiplet with an  $\td{\alg}$-valued $D=3$ super Yang-Mills multiplet yields a $D=3$ supergravity mulitplet with fields valued in $\alg\otimes\td{\alg}$, making a magic square of U-dualities appear rather natural.

Let us now review in some more detail these constructions. The Lagrangian for $(n+2)$-dimensional $\mathcal{N}=1$ super Yang-Mills with gauge group $G$ over the division algebra $\alg_n$ is given \cite{Anastasiou:2013cya} by
\begin{equation}\label{10-d action}
\mathcal{L}(\alg_n)=-\frac{1}{4}F_{\mu\nu}^AF^{A\mu\nu}-\text{Re}(i\lambda^{\dagger A}\sigmabar^\mu D_\mu\lambda^A),~~~~~\lambda\in\alg_n^2,
\end{equation}
where the covariant derivative and field strength are given by the usual expressions
\be
\begin{split}
D_\mu\lambda^A&=\partial_\mu\lambda^A+gf_{BC}{}^A A_\mu^B\lambda^C,\\
F^A_{\mu\nu}&=\partial_\mu A^A_\nu-\partial_\nu A_\mu^A+gf_{BC}{}^AA^B_\mu A^C_\nu,
\end{split}
\ee
with $A=0,\dots,\dim[G]$. The $\{\sigma^\mu\}$ are a basis for $\alg_n$-valued Hermitian matrices - the straightforward generalisation of the usual complex Pauli matrices \cite{Schray:1994ur, Evans:1994cn, Anastasiou:2013cya} to all four normed division algebras, satisfying the usual Clifford algebra relations. We can use these to write the supersymmetry transformations:
\be\label{SUSYTRANS}
\delta A_\mu^A=\text{Re}(i\lambda^{\dagger A}\sigmabar_\mu \epsilon),\hspace{0.5cm}
\delta\lambda^A=\frac{1}{4}F_{\mu\nu}^A\sigma^\mu(\sigmabar^\nu\epsilon).
\ee
Note, since the octonions are non-associative the ordering of the parentheses is important. Moreover, the components $\lambda^{Aa}$ are anti-commuting; we are dealing with the algebra of octonions defined over the Grassmanns and we cannot rely on the usual spinor identities to hold automatically. However, everything goes through, thanks principally to alternativity.

By dimensionally reducing these theories using the Dixon-halving techniques of \cite{Anastasiou:2013cya}, we arrive at the Lagrangian for super Yang-Mills in $D=n+2$ with $\N$ supersymmetries written over the division algebra $\alg_{n\N}$. The division algebra associated with spacetime $\alg_n$ is viewed as a  subalgebra of $\alg_{n\N}$. The resulting Lagrangian is:
\be
\begin{split}
\mathcal{L}\left(\alg_n,\alg_{n\N}\right)=&-\frac{1}{4}F^A_{\mu\nu}F^{A\mu\nu}-\frac{1}{2}\langle D_\mu\phi^{A}|D^\mu\phi^A\rangle-\text{Re}(i\lambda^{\dagger A}\sigmabar^\mu D_\mu\lambda^A)  \label{MASTER}\\ 
&-gf_{BC}{}^A\text{Re}\left(i{\lambda}^{\dagger A}\varepsilon\phi^B\lambda^C\right)-\frac{1}{16}g^2f_{BC}{}^Af_{DE}{}^A\langle\phi^B|\phi^D\rangle\langle\phi^C|\phi^E\rangle,
\end{split}
\ee
where $\lambda\in\alg_{n\N}^2$ (so we have $\N$ spacetime spinors, each valued in $\alg_{n}^2$) and $\phi$ is a scalar field taking values in $\phi\in\alg_n^\complement$,  the subspace of $\alg_{n\N}$ orthogonal to the $\alg_n$ subalgebra. The $\{\sigmabar^\mu\}$ are still a basis for $\alg_n$-valued Hermitian matrices, again, with $\alg_n$ viewed as a division subalgebra of $\alg_{n\N}$.  As noted in \cite{Anastasiou:2013cya}, the overall (spacetime little group plus internal) symmetry of the $\N=1$ theory in $D=n+2$ dimensions is given by the triality algebra, $\mf{tri}(\alg_n)$. If we dimensionally reduce these theories we obtain super Yang-Mills with $\N$ supersymmetries whose overall symmetries are given by,
\be
\mf{sym}(\alg_n,\alg_{n\N}):=\big\{(A,B,C)\in\mathfrak{tri}(\alg_{n\N})| [A, \mf{so}(\alg_n)_{ST}]=0, ~~\forall A\notin \mf{so}(\alg_n)_{ST} \big\},
\ee
where $\mf{so}(\alg_n)_{ST}$ is the subalgbra of $\mathfrak{so}(\alg_{n\N})$ that acts as orthogonal transformations on $\alg_n\subseteq\alg_{n\N}$. The division algebras used in each dimension and the corresponding $\mf{sym}$ algebras are summarised in  \autoref{YANG}.
 \begin{table}[ht]
 \begin{center}
\begin{tabular}{c|ccccccc}
\hline
\hline
 $\alg_n\backslash\alg_{n\N}$ & $\hspace{1.4cm}\Oct\hspace{1.4cm}$ & $\hspace{1.4cm}\Q\hspace{1.4cm}$  & $\hspace{1.4cm}\C\hspace{1.4cm}$  & $\hspace{1.2cm}\R\hspace{1.0cm}$ & \\
 \hline
 \\
 
   $\Oct$ & $\mf{so}(8)_{ST}$ \\ \\
$\Q$ & $\mf{so}(4)_{ST}\oplus\mf{sp}(1)\oplus\mf{sp}(1)$ & $\mf{so}(4)_{ST}\oplus\mf{sp}(1)$
\\ \\
$\C$ & $\mf{so}(2)_{ST}\oplus\mf{su}(4)$ & $\mf{so}(2)_{ST}\oplus\mf{sp}(1)\oplus\mf{so}(2)$
& $\mf{so}(2)_{ST}\oplus\mf{so}(2)$ \\ \\
$\R$ & $\mf{so}(8)$ & $\mf{so}(4)\oplus\mf{sp}(1)$ & $\mf{so}(2)\oplus\mf{so}(2)$ & $\emptyset$
\\ \\
   \hline
   \hline
\end{tabular}
\caption{\footnotesize{A table of algebras: $\mf{sym}(\alg_n,\alg_{n\N})$. This lets us read off the spacetime and internal symmetries in each Yang-Mills theory. For example, one can see the familiar R-symmetries in $D=4$: U(1), U(2) and SU(4) for $\N=1,2,4$, respectively. Note that the symmetries in $D=3$ are entirely internal and that they include the R-symmetry as a subgroup (these are actually the symmetries of the theories after dualising the vector to a scalar)}.\label{YANG}}
 \end{center}
\end{table}

Let us take $D=3$ as a concrete example. The   $\mathcal{N}=8$  Lagrangian is given by
\be
\begin{split}
\mathcal{L}=&-\tfrac{1}{4}F^A_{\mu\nu}F^{A\mu\nu}-\tfrac{1}{2}D_\mu\phi_i^{A}D^\mu\phi_i^A+i\bar{\lambda}_a^{ A}\gamma^\mu D_\mu\lambda_a^{A}  \\ 
& -\tfrac{1}{4}g^2f_{BC}{}^Af_{DE}{}^A\phi_i^B\phi_i^D\phi^C_j\phi^E_j  \\&-gf_{BC}{}^A\phi^B_i \bar{\lambda}^{ Aa}\Gamma^i_{ab} \lambda^{Cb},
\end{split}
\ee
where $\Gamma^i_{ab}$, $i=1,\ldots,7$, $a,b=0,\ldots,7$, belongs to the SO(7) Clifford algebra. The key observation is that this gamma matrix can be represented by the octonionic structure constants,
\be
\Gamma^i_{ab}=i(\delta_{bi}\delta_{a0}-\delta_{b0}\delta_{ai}+C_{iab}),
\ee
which allows us to rewrite the action over octonionic fields. If we replace $\Oct$ with a general division algebra $\alg$, the result is $\mathcal{N}=1,2,4,8$ over $\R,\C,\Q,\Oct$:
\be
\begin{split}
\mathcal{L}=&- \tfrac{1}{4}F^A_{\mu\nu}F^{A\mu\nu}-\tfrac{1}{2}D_\mu\phi^{*A}D^\mu\phi^A+i\bar{\lambda}^{ A}\gamma^\mu D_\mu\lambda^A  \\ 
&-\tfrac{1}{4}g^2f_{BC}{}^Af_{DE}{}^A\langle\phi^B|\phi^D\rangle\langle\phi^C|\phi^E\rangle \hspace{.95cm}  \\
 &+\tfrac{i}{2}gf_{BC}{}^A\left((\bar{\lambda}^{A}\phi^B)\lambda^C-\bar{\lambda}^{A}(\phi^{*B}\lambda^C)\right),
\end{split}
\ee
where $\phi=\phi^i e_i$ is an Im$\mathds{A}$-valued scalar field, $\lambda=\lambda^a e_a$ is an $\mathds{A}$-valued two-component spinor and $\bar{\lambda}=\bar{\lambda}^ae_a^*$. 

The supersymmetry transformations in this language are given by
\begin{eqnarray}
\delta\lambda^A&&=\frac{1}{2}(F^{A\mu\nu}+\varepsilon^{\mu\nu\rho}D_\rho\phi^A)\sigma_{\mu\nu}\epsilon-\frac{1}{4}gf_{BC}{}^A\phi^B(\phi^C\epsilon), \nonumber\\
\delta A_\mu^A&&=\frac{i}{2}(\bar{\epsilon}\gamma_\mu\lambda^A-\bar{\lambda}^{ A}\gamma_\mu\epsilon),\\
\delta \phi^A&&= \frac{i}{2}e_i[(\bar{\epsilon} e_i)\lambda^A-\bar{\lambda}^{ A}(e_i\epsilon)],\nonumber
\end{eqnarray}
where $\epsilon$ is an $\mathds{A}$-valued two-component spinor and $\sigma_{\mu\nu}$ are the generators of $\SL(2,\R)\cong\Spin(1,2)$.  The form of the first term in the $\lambda^A$ transformation also highlights the vector's status as the missing real part of the Im$\mathds{A}$-valued scalar field. Indeed, in the free $g=0$ theory one may dualise the vector to a scalar to obtain a full $\mathds{A}$-valued field.

Now consider the product of two division algebraic multiplets:
\begin{enumerate}
\item A left $\N=\dim \alg$ multiplet 
\be
\{A_\mu \in \text{Re}\alg,\quad \phi \in \text{Im}\alg,\quad \lambda\in\alg\}
\ee 
\item A right $\tN=\dim \td{\alg}$ multiplet
\be
\{{\tA}_\nu \in \text{Re}\tilde{\alg}, \quad\tilde{\phi} \in \text{Im}\tilde{\alg}, \quad\tilde{\lambda}\in\tilde{\alg}\}
\ee
\end{enumerate}
We obtain the field content of an $(\N+\tN)$-extended supergravity theory valued in both $\tilde{\alg}$ and $\tilde{\alg}$:
\be\label{eq:sugrafields}
g_{\mu\nu}  \in \R, \quad
\Psi_{\mu} \in  \begin{pmatrix} \alg\\ \tilde{\alg}\end{pmatrix}, \quad
\varphi, \chi  \in  \begin{pmatrix} \alg\otimes \tilde{\alg}\\ \alg\otimes \tilde{\alg}\end{pmatrix}.
\ee
The $\R$-valued graviton and $\alg\oplus \tilde{\alg}$-valued gravitino carry no degrees of freedom. The $(\alg\otimes\tilde{\alg})^2$-valued scalar and Majorana spinor each have $2(\dim \alg\times\dim\tilde{\alg})$ degrees of freedom.  

\begin{figure}[h]
\centering
\includegraphics[scale=0.3]{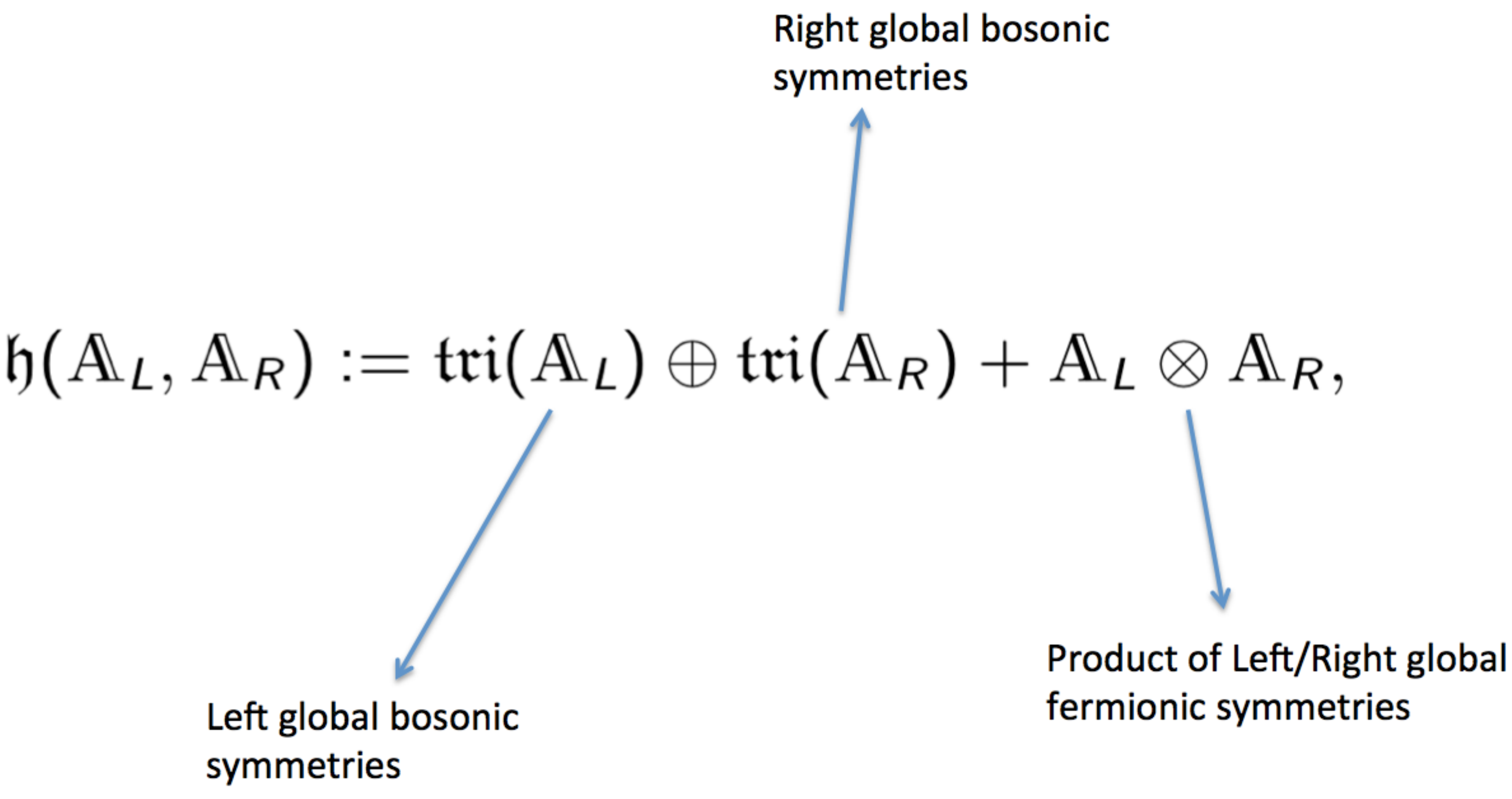}
\caption{The $\mathcal{H}$ algebra in terms of the left/right super Yang-Mills theories in a division algebraic language.}\label{halgpic}
\end{figure}

The $\mathcal{H}$ algebra then follows immediately  in this division algebraic language. The left and right factors each come with a commuting copy of the triality algebra,
$\mf{tri}(\alg)\oplus\mf{tri}(\tilde{\alg})$.
However,    the $\alg\otimes\tilde{\alg}$ doublets in  \eqref{eq:sugrafields} form irreducible representations of R-symmetry. The corresponding   generators  must themselves transform under $\mf{tri}(\alg)\oplus\mf{tri}(\tilde{\alg})$  consistently,  implying they are elements  of $\alg\otimes\tilde{\alg}$. This follows, formally, from the left/right supersymmetries. The conventional infinitesimal supersymmetry variation of the $left\otimes right$   states correctly gives the infinitesimal supersymmetry variation on the corresponding supergravity states \cite{Siegel:1995px, Bianchi:2008pu, Anastasiou:2014qba}. Seeking, instead,  internal \emph{bosonic} transformations on the supergavity multiplet suggests starting from the rather unconventional tensor product of the left and right supercharges, $Q\otimes \tilde{Q}$.  See \autoref{halgpic}. 
This  follows, at least formally,  from the observation
$Q\otimes \tilde{Q}\in \alg \otimes \tilde{\alg}$, 
where we are explicitly suppressing the spacetime representation space. Note,  these are ``pseudo-supersymmetry'' transformations since they do not change the mass dimension of the component fields. For an explicit construction see \cite{Anastasiou:2015vba}. In summary, we have in total:
\be\label{h}
\mf{h}(\alg,  \tilde{\alg}) :=\mf{tri}(\alg)\oplus\mf{tri}(\tilde{\alg})+\alg \otimes \tilde{\alg}.
\ee
This Lie algebra, see \cite{Anastasiou:2013hba} for the commutators, yields the maximal compact subalgebras of the corresponding U-dualities, given in 
\autoref{tab:ms3}.
  \begin{table}[ht]
 \begin{center}
\begin{tabular}{c|ccccccc}
\hline
\hline
 $\alg_L\backslash\alg_R$ && $\R$ & $\C$  & $\Q$  & $\Oct$ & \\
 \hline
 \\
 
   $\R$ && $\mf{so}(2)$ & $\mf{so}(3)\times \mf{so}(2)$   & $\mf{so}(5)\times \mf{so}(3)$   & $\mf{so}(9)$   \\
  $\C$ && $\mf{so}(3)\times \mf{so}(2)$ & $[\mf{so}(3)\times \mf{so}(2)]^2$   & $\mf{so}(6)\times \mf{so}(3)\times \mf{so}(2)$   & $\mf{so}(10)\times \mf{so}(2)$   \\
  $\Q$ && $\mf{so}(5)\times \mf{so}(3)$ & $\mf{so}(6)\times \mf{so}(3)\times \mf{so}(2)$   &$\mf{so}(8)\times \mf{so}(4)$  & $\mf{so}(12)\times \mf{so}(3)$   \\
   $\Oct$ && $\mf{so}(9)$ & $\mf{so}(10)\times \mf{so}(2)$   & $\mf{so}(12)\times \mf{so}(3)$   & $\mf{so}(16)$   \\
   \\
   \hline
   \hline
\end{tabular}
\caption[Magic square of required real forms.]{Magic square of maximal compact subalgebras.  \label{tab:ms3}}
 \end{center}
\end{table}

\begin{figure}[h]
\centering
\includegraphics[scale=0.3]{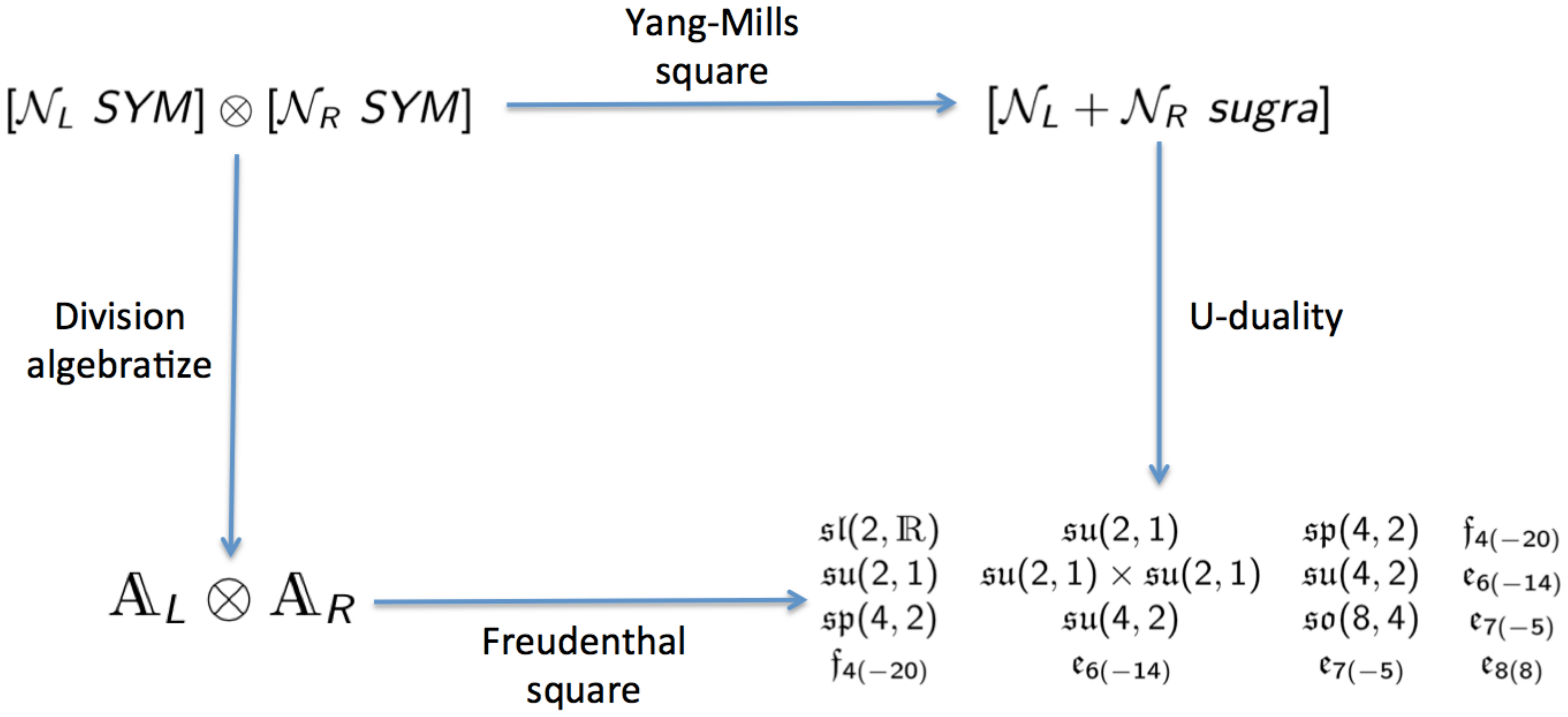}
\caption{All roads lead to the magic square.}\label{allroads}
\end{figure}

The U-dualities $\mathcal{G}$ are realised non-linearly on the scalars, which parametrise the symmetric spaces $\mathcal{G}/\mathcal{H}$.    This can be  understood using the  identity relating $(\alg\otimes\tilde{\alg})^2$ to  $\mathcal{G}/\mathcal{H}$, 
\be
(\alg\otimes\tilde{\alg})\mathds{P}^2 \cong \mathcal{G}/\mathcal{H}.
\ee
The scalar fields may be regarded as points in division-algebraic projective planes. The tangent space $T_{p}(\mathcal{G/H})\cong\mathfrak{p}=\mathfrak{g}\ominus\mathfrak{h}$ implies the scalars carry the $\mathfrak{p}$-representation of $\mathcal{H}$. The tangent space at any point of $(\alg\otimes\tilde{\alg})\mathds{P}^2$ is just $(\alg\otimes\tilde{\alg})^2$, the required representation space  of $\mathcal{H}$. Since $\mathcal{G}/\mathcal{H}$ is a symmetric space,  the  U-duality Lie algebra is given by adjoining the scalar representation space $(\alg\otimes\tilde{\alg})^2$ to \autoref{halgpic},
\be\label{msms}
\mf{m}(\alg, \tilde{\alg}):=\underbrace{\mf{tri}(\alg)\oplus \mf{tri}(\tilde{\alg})+(\alg\otimes\tilde{\alg})}_{\mf{h}(\alg, \tilde{\alg})}+\underbrace{(\alg\otimes\tilde{\alg})^2}_{\text{``scalars''}}.
\ee
This has a $\Z_2\times\Z_2$ graded Lie algebra structure uniquely determined by the left/right super Yang-Mills factors and yields precisely the   magic square. See \cite{Anastasiou:2013hba} for a full account of the commutation relations. The triality construction  described in \cite{Barton:2003} is isomorphic to \eqref{msms} as a vector space, but has a different Lie algebra structure, as reflected in the distinct real forms appearing in each case. In conclusion, the  product of division algebras and super Yang-Mills theories both lead to the magic square, as depicted in \autoref{allroads}.

For $D=n+2$, we begin with a pair of Yang-Mills theories with $\N$ and $\tN$ supersymmetries written over the division algebras $\alg_{n\N}$ and $\alg_{n\tN}$, respectively, as described by \eqref{MASTER}. In terms of spacetime little group representations we may then write all the bosons of the left (right) theory as a single element $b\in \alg_{n\N}$ ($\td{b}\in \alg_{n\tN}$), and similarly for the fermions $f\in \alg_{n\N}$ ($\td{f}\in \alg_{n\tN}$). After tensoring we arrange the resulting supergravity fields into a bosonic doublet and a fermionic doublet, 
\be\label{eq:sugrafieldsonshellD}
B= \begin{pmatrix} b\otimes \td{b}\\ f\otimes \td{f} \end{pmatrix},\quad F  =  \begin{pmatrix} b\otimes \td{f}\\f\otimes \td{b}\end{pmatrix},
\ee
just as we did in $D=3$. The  algebra \eqref{h}  acts naturally  on these doublets. However,  a diagonal $\mf{so}(\alg_n)_{ST}$ subalgebra of this corresponds to spacetime transformations, so we must restrict $\mf{h}(\alg_{n\N},\alg_{n\tN})$ to the subalgebra that commutes with $\mf{so}(\alg_n)_{ST}$. Heuristically, we identify a diagonal spacetime subalgebra $\alg_n$ in  $\alg_{n\N}\otimes\alg_{n\tN}$ and require that it is preserved by the global isometries, which picks out a subset in $\mf{Isom}((\alg_{n\N}\otimes\alg_{n\tN})\mathds{P}^2)$. Imposing this condition selects the U-duality   algebra of the $D=n+2,$ $(\N+\tN)$-extended supergravity theory obtained by tensoring  $\N$ and $\tN$  super Yang-Mills theories. The Lie algebras are given by the \emph{magic pyramid formula}:
\be\label{eq:pyramid}
\mathfrak{MPyr}(\alg_{n}, \alg_{n\N}, \alg_{n\tN}):=\left\{u\in\mf{m}(\alg_{n\N}, \alg_{n\tN})-\mf{so}(\alg_n)_{ST}\Big|[u,\mf{so}(\alg_n)_{ST}]=0\right\}.
\ee
\begin{figure}
\begin{center}
\includegraphics[scale=0.12]{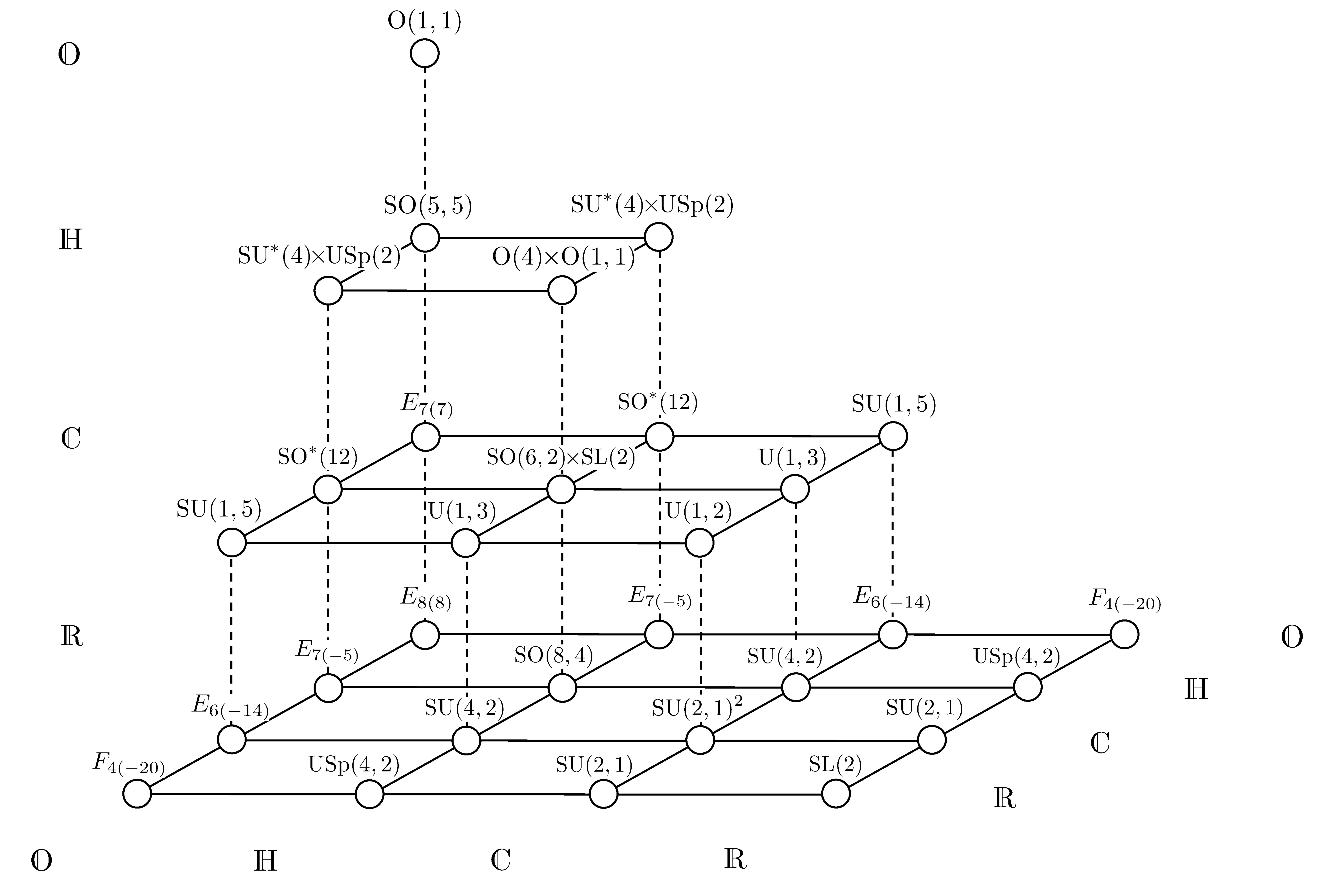}
\caption{\footnotesize{A magic pyramid of supergravities. The vertical axis labels the spacetime division algebra $\alg_n$, while the horizontal axes label the algebras associated with the number of supersymmetries $\alg_{n \mathcal{N}}$ and $\alg_{n\tN}$.}}\label{pyramid}
\end{center}
\end{figure}
The terminology is made clear by the pyramid of corresponding U-dualities groups presented in \autoref{pyramid}. The base of the  pyramid in $D=3$ is the  $4 \times4$ Freudenthal magic square,  while the higher levels are comprised of a $3 \times 3$ square in $D=4$, a $2 \times 2$ square in $D=6$ and Type II supergravity at the apex in $D=10$. Note, in \cite{Julia:1980gr}  the oxidation of $\mathcal{N}$-extended  $D=3$ dimensional supergravity theories was shown to generate a partially symmetric ``trapezoid'' of non-compact global symmetries for $D=3,4,\ldots 11$ and $0, 2^0, 2^1,\ldots 2^7$ supercharges. A subset of algebras in the trapezoid with $D=3,4,5$ and $2^5, 2^6, 2^7$ supercharges  matches  the $D=3,4,5$ and $\alg=\C, \Q, \Oct$ exterior wall of the pyramid of \autoref{pyramid}.

\begin{figure}
\begin{center}
\includegraphics[scale=0.5]{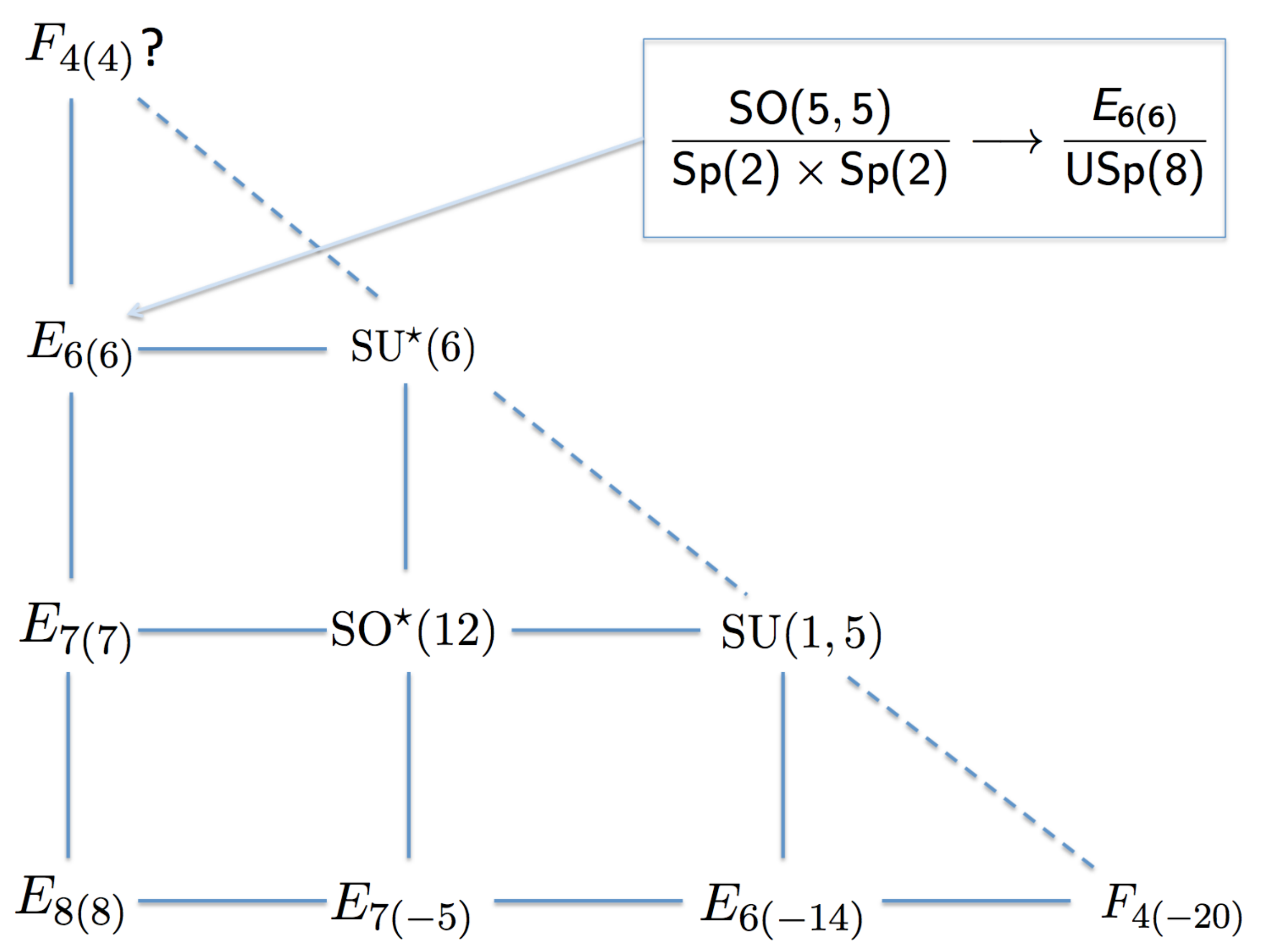}
\caption{The magic faces of the conformal pyramid. We have highlighted the the maxiamally supersymmetric case in $D=6$. The $\N=(2,2)$ supergravity theory of the magic pyramid with U-Duality $\SO(5,5)$ is replaced by the $\N=(4, 0)$ superconfromal theory, proposed by Hull \cite{Hull:2000zn, Hull:2000rr,  Hull:2000ih}, with U-duality $E_{6(6)}$. As a consequence, the outer faces of the conformal pyramid are diagonal slices of the Freudenthal magic square, upto the $F_{4(4)}$ slot, which currently has does not have a  theory associated to it.}\label{magicface}
\end{center}
\end{figure}

Let us conclude with some comments on the product of  theories other than super Yang-Mills. Particularly interesting examples are provided by the superconformal  multiplets in $D=3,4,6$. In a manner directly analogous to the magic pyramid the tensor product of left and right superconformal theories yields the ``conformal pyramid'', described in \cite{Anastasiou:2013hba}. It has the remarkable property that its faces are also given by the  Freudenthal magic square, as depicted in \autoref{magicface}. In particular, ascending up the maximal spine one encounters  the famous exceptional sequence $E_{8(8)}, E_{7(7)}, E_{6(6)}$, but where $E_{6(6)}$ belongs to the  $D=6, (4, 0)$ theory proposed by Hull as the superconformal limit of M-theory compactified on a 6-torus \cite{Hull:2000zn, Hull:2000rr, Hull:2000ih}.  This pattern  suggests the existence of some highly exotic $D=10$ theory with $F_{4(4)}$ U-duality group. The existence of such a theory would be more than a little  surprising and there is a (slightly) more conventional interpretation of the conformal pyramid, including its $F_{4(4)}$ tip, but for theories in $D=3,4,5,6$, as described \cite{Anastasiou:2013hba}.

The product of conformal theories in the context of amplitudes has been considered previously in, for example,  \cite{Huang:2010rn, Czech:2011dk, Chiodaroli:2011pp, Bargheer:2012gv, Huang:2012wr}. In particular, the maximally supersymmetric  $D=3$, $\N=8$ Bagger-Lambert-Gustavsson (BLG) Chern-Simons-matter theory \cite{Bagger:2006sk, Gustavsson:2007vu,Bagger:2007jr} has   been shown to enjoy a colour-kinematic duality reflecting its three-algebra structure \cite{Bargheer:2012gv}. The ``square'' of BLG amplitudes yields those  of $\N=16$ supergravity. Since $\N=16$ supergravity is the unique theory with 32 supercharges in three dimensions it is also the ``square'' of the $\N=8$ Yang-Mills theory. The square of the amplitudes  in both cases agree, despite their distinct structures \cite{Huang:2012wr}.

In $D=6$ one might expect relations between the ``square'' of the  $\N=(2,0)$ tensor multiplet  and the   $\N=(4, 0)$ theory proposed by Hull \cite{Hull:2000zn, Hull:2000rr,  Hull:2000ih}, as discussed in \cite{Chiodaroli:2011pp}. Of course, amplitudes are generically not well-defined  in these cases, but one can make some precise statements in terms of the  tree-level $S$-matrix in particular regimes, as discussed in \cite{Huang:2010rn, Czech:2011dk}. For example, in the absence of additional degrees of freedom all tree-level amplitudes of the $(2,0)$ tensor multiplet vanish \cite{Huang:2010rn}. The $D=5, \N=4$ super Yang-Mills theory squares to give the amplitudes of $D=5, \N=8$ supergravity. However, being non-renormalisable it ought to be regarded as  a  superconformal   $D=6, \N=(2,0)$   theory compactified on a circle of radius $R=g_{YM}^{2}/4\pi^2$.   At linearised level  Hull's $(4, 0)$ theory follows from the square the $(2, 0)$ theory \cite{Borsten:2015toappear} and   gives  $\N=8, D=5$ supergravity when compactified on a circle \cite{Hull:2000rr}. From this perspective the $(2, 0) \times (2, 0) = (4, 0)$ identity constitutes  an, as yet ill-defined,  M-theory up-lift of the  maximally supersymmetric $D=5$ squaring relation.  

\section*{Acknowledgments}

MJD is grateful to the organisers of the Arnowitt Memorial Symposium for the invitation to contribute.  Based on work done in collaboration with A.~Anastasiou,  M.~J.~Hughes and
S.~Nagy. We are very grateful for  their numerous and essential contributions.   We would also like to thank A.~Marrani for his insights on the symmetries of supergravity. The work of LB was supported by a Sch\"{o}dinger Fellowship. The work of MJD is supported by the STFC under rolling grant ST/G000743/1.


\begin{thebibliography}{100}

\bibitem{1956PhRv..101.1597U}
R.~{Utiyama}, \href{http://dx.doi.org/10.1103/PhysRev.101.1597}{``{Invariant
  Theoretical Interpretation of Interaction},''{\em Physical Review} {\bf 101}
  (Mar., 1956)  1597--1607}.

\bibitem{1961JMP.....2..212K}
T.~W.~B. {Kibble}, \href{http://dx.doi.org/10.1063/1.1703702}{``{Lorentz
  Invariance and the Gravitational Field},''{\em Journal of Mathematical
  Physics} {\bf 2} (Mar., 1961)  212--221}.

\bibitem{1977PhRvL..38..739M}
S.~W. {MacDowell} and F.~{Mansouri},
  \href{http://dx.doi.org/10.1103/PhysRevLett.38.739}{``{Unified geometric
  theory of gravity and supergravity},''{\em Physical Review Letters} {\bf 38}
  (Apr., 1977)  739--742}.

\bibitem{1977NuPhB.129...39C}
A.~H. {Chamseddine} and P.~C. {West},
  \href{http://dx.doi.org/10.1016/0550-3213(77)90018-9}{``{Supergravity as a
  gauge theory of supersymmetry},''{\em Nuclear Physics B} {\bf 129} (Oct.,
  1977)  39--44}.

\bibitem{1979JPhA...12L.205S}
K.~S. {Stelle} and P.~C. {West},
  \href{http://dx.doi.org/10.1088/0305-4470/12/8/003}{``{de Sitter gauge
  invariance and the geometry of the Einstein-Cartan theory},''{\em Journal of
  Physics A Mathematical General} {\bf 12} (Aug., 1979)  L205--L210}.

\bibitem{hooft1994dimensional}
G.~Hooft, ``Dimensional reduction in quantum gravity,'' {\em Salamfestschrift:
  a collection of talks} {\bf 4} (1994) no.~A, 1--13,
  \href{http://arxiv.org/abs/gr-qc/9310026}{{\tt gr-qc/9310026}}.

\bibitem{Susskind:1994vu}
L.~Susskind, ``{The World as a hologram},''
  \href{http://dx.doi.org/10.1063/1.531249}{{\em J.Math.Phys.} {\bf 36} (1995)
  6377--6396},
\href{http://arxiv.org/abs/hep-th/9409089}{{\tt arXiv:hep-th/9409089
  [hep-th]}}.

\bibitem{Maldacena:1997re}
J.~M. Maldacena, ``{The large N limit of superconformal field theories and
  supergravity},'' \href{http://dx.doi.org/10.1023/A:1026654312961}{{\em Adv.
  Theor. Math. Phys.} {\bf 2} (1998)  231--252},
\href{http://arxiv.org/abs/hep-th/9711200}{{\tt arXiv:hep-th/9711200}}.

\bibitem{Witten:1998qj}
E.~Witten, ``{Anti-de Sitter space and holography},'' {\em
  Adv.Theor.Math.Phys.} {\bf 2} (1998)  253--291,
\href{http://arxiv.org/abs/hep-th/9802150}{{\tt arXiv:hep-th/9802150
  [hep-th]}}.

\bibitem{Gubser:1998bc}
S.~Gubser, I.~R. Klebanov, and A.~M. Polyakov, ``{Gauge theory correlators from
  noncritical string theory},''
  \href{http://dx.doi.org/10.1016/S0370-2693(98)00377-3}{{\em Phys.Lett.} {\bf
  B428} (1998)  105--114},
\href{http://arxiv.org/abs/hep-th/9802109}{{\tt arXiv:hep-th/9802109
  [hep-th]}}.

\bibitem{Kawai:1985xq}
H.~Kawai, D.~Lewellen, and S.~Tye, ``{A Relation Between Tree Amplitudes of
  Closed and Open Strings},''
\href{http://dx.doi.org/10.1016/0550-3213(86)90362-7}{{\em Nucl.Phys.} {\bf
  B269} (1986)  1}.

\bibitem{Bern:2008qj}
Z.~Bern, J.~Carrasco, and H.~Johansson, ``{New Relations for Gauge-Theory
  Amplitudes},'' \href{http://dx.doi.org/10.1103/PhysRevD.78.085011}{{\em
  Phys.Rev.} {\bf D78} (2008)  085011},
\href{http://arxiv.org/abs/0805.3993}{{\tt arXiv:0805.3993 [hep-ph]}}.

\bibitem{Bern:2010ue}
Z.~Bern, J.~J.~M. Carrasco, and H.~Johansson, ``{Perturbative Quantum Gravity
  as a Double Copy of Gauge Theory},''
  \href{http://dx.doi.org/10.1103/PhysRevLett.105.061602}{{\em Phys.Rev.Lett.}
  {\bf 105} (2010)  061602},
\href{http://arxiv.org/abs/1004.0476}{{\tt arXiv:1004.0476 [hep-th]}}.

\bibitem{Elvang:2013cua}
H.~Elvang and Y.-t. Huang, ``{Scattering Amplitudes},''
\href{http://arxiv.org/abs/1308.1697}{{\tt arXiv:1308.1697 [hep-th]}}.

\bibitem{Bern:2010yg}
Z.~Bern, T.~Dennen, Y.-t. Huang, and M.~Kiermaier, ``{Gravity as the Square of
  Gauge Theory},'' \href{http://dx.doi.org/10.1103/PhysRevD.82.065003}{{\em
  Phys.Rev.} {\bf D82} (2010)  065003},
\href{http://arxiv.org/abs/1004.0693}{{\tt arXiv:1004.0693 [hep-th]}}.

\bibitem{Carrasco:2015iwa}
J.~J.~M. Carrasco, ``{Gauge and Gravity Amplitude Relations},''
\href{http://arxiv.org/abs/1506.00974}{{\tt arXiv:1506.00974 [hep-th]}}.

\bibitem{Bern:2009kd}
Z.~Bern, J.~J. Carrasco, L.~J. Dixon, H.~Johansson, and R.~Roiban, ``{The
  Ultraviolet Behavior of N=8 Supergravity at Four Loops},''
  \href{http://dx.doi.org/10.1103/PhysRevLett.103.081301}{{\em Phys. Rev.
  Lett.} {\bf 103} (2009)  081301},
\href{http://arxiv.org/abs/0905.2326}{{\tt arXiv:0905.2326 [hep-th]}}.

\bibitem{Bern:2014sna}
Z.~Bern, S.~Davies, and T.~Dennen, ``{Enhanced ultraviolet cancellations in
  $\mathcal N=5$ supergravity at four loops},''
  \href{http://dx.doi.org/10.1103/PhysRevD.90.105011}{{\em Phys.Rev.} {\bf D90}
  (2014) no.~10, 105011},
\href{http://arxiv.org/abs/1409.3089}{{\tt arXiv:1409.3089 [hep-th]}}.

\bibitem{Monteiro:2011pc}
R.~Monteiro and D.~O'Connell, ``{The Kinematic Algebra From the Self-Dual
  Sector},'' \href{http://dx.doi.org/10.1007/JHEP07(2011)007}{{\em JHEP} {\bf
  1107} (2011)  007},
\href{http://arxiv.org/abs/1105.2565}{{\tt arXiv:1105.2565 [hep-th]}}.

\bibitem{Monteiro:2013rya}
R.~Monteiro and D.~O'Connell, ``{The Kinematic Algebras from the Scattering
  Equations},'' \href{http://dx.doi.org/10.1007/JHEP03(2014)110}{{\em JHEP}
  {\bf 1403} (2014)  110},
\href{http://arxiv.org/abs/1311.1151}{{\tt arXiv:1311.1151 [hep-th]}}.

\bibitem{BjerrumBohr:2009rd}
N.~E.~J. Bjerrum-Bohr, P.~H. Damgaard, and P.~Vanhove, ``{Minimal Basis for
  Gauge Theory Amplitudes},''
  \href{http://dx.doi.org/10.1103/PhysRevLett.103.161602}{{\em Phys. Rev.
  Lett.} {\bf 103} (2009)  161602},
\href{http://arxiv.org/abs/0907.1425}{{\tt arXiv:0907.1425 [hep-th]}}.

\bibitem{Stieberger:2009hq}
S.~Stieberger, ``{Open \& Closed vs. Pure Open String Disk Amplitudes},''
\href{http://arxiv.org/abs/0907.2211}{{\tt arXiv:0907.2211 [hep-th]}}.

\bibitem{Tye:2010dd}
S.~H. Henry~Tye and Y.~Zhang, ``{Dual Identities inside the Gluon and the
  Graviton Scattering Amplitudes},''
  \href{http://dx.doi.org/10.1007/JHEP06(2010)071,
  10.1007/JHEP04(2011)114}{{\em JHEP} {\bf 06} (2010)  071},
  \href{http://arxiv.org/abs/1003.1732}{{\tt arXiv:1003.1732 [hep-th]}}.
[Erratum: JHEP04,114(2011)].

\bibitem{Mafra:2011kj}
C.~R. Mafra, O.~Schlotterer, and S.~Stieberger, ``{Explicit BCJ Numerators from
  Pure Spinors},'' \href{http://dx.doi.org/10.1007/JHEP07(2011)092}{{\em JHEP}
  {\bf 07} (2011)  092},
\href{http://arxiv.org/abs/1104.5224}{{\tt arXiv:1104.5224 [hep-th]}}.

\bibitem{Mafra:2015vca}
C.~R. Mafra and O.~Schlotterer, ``{Berends-Giele recursions and the BCJ duality
  in superspace and components},''
\href{http://arxiv.org/abs/1510.08846}{{\tt arXiv:1510.08846 [hep-th]}}.

\bibitem{Mafra:2014gja}
C.~R. Mafra and O.~Schlotterer, ``{Towards one-loop SYM amplitudes from the
  pure spinor BRST cohomology},''
  \href{http://dx.doi.org/10.1002/prop.201400076}{{\em Fortsch. Phys.} {\bf 63}
  (2015) no.~2, 105--131},
\href{http://arxiv.org/abs/1410.0668}{{\tt arXiv:1410.0668 [hep-th]}}.

\bibitem{Mafra:2015mja}
C.~R. Mafra and O.~Schlotterer, ``{Two-loop five-point amplitudes of super
  Yang-Mills and supergravity in pure spinor superspace},''
  \href{http://dx.doi.org/10.1007/JHEP10(2015)124}{{\em JHEP} {\bf 10} (2015)
  124},
\href{http://arxiv.org/abs/1505.02746}{{\tt arXiv:1505.02746 [hep-th]}}.

\bibitem{He:2015wgf}
S.~He, R.~Monteiro, and O.~Schlotterer, ``{String-inspired BCJ numerators for
  one-loop MHV amplitudes},''
  \href{http://dx.doi.org/10.1007/JHEP01(2016)171}{{\em JHEP} {\bf 01} (2016)
  171},
\href{http://arxiv.org/abs/1507.06288}{{\tt arXiv:1507.06288 [hep-th]}}.

\bibitem{Howe:1988qz}
P.~S. Howe and K.~Stelle, ``{The Ultraviolet Properties of Supersymmetric Field
  Theories},''
\href{http://dx.doi.org/10.1142/S0217751X89000753}{{\em Int.J.Mod.Phys.} {\bf
  A4} (1989)  1871}.

\bibitem{Deser:1977nt}
S.~Deser, J.~H. Kay, and K.~S. Stelle, ``{Renormalizability Properties of
  Supergravity},'' \href{http://dx.doi.org/10.1103/PhysRevLett.38.527}{{\em
  Phys. Rev. Lett.} {\bf 38} (1977)  527},
\href{http://arxiv.org/abs/1506.03757}{{\tt arXiv:1506.03757 [hep-th]}}.

\bibitem{Green:2010sp}
M.~B. Green, J.~G. Russo, and P.~Vanhove, ``{String theory dualities and
  supergravity divergences},''
  \href{http://dx.doi.org/10.1007/JHEP06(2010)075}{{\em JHEP} {\bf 1006} (2010)
   075},
\href{http://arxiv.org/abs/1002.3805}{{\tt arXiv:1002.3805 [hep-th]}}.

\bibitem{Bossard:2010bd}
G.~Bossard, P.~Howe, and K.~Stelle, ``{On duality symmetries of supergravity
  invariants},'' \href{http://dx.doi.org/10.1007/JHEP01(2011)020}{{\em JHEP}
  {\bf 1101} (2011)  020},
\href{http://arxiv.org/abs/1009.0743}{{\tt arXiv:1009.0743 [hep-th]}}.

\bibitem{Beisert:2010jx}
N.~Beisert, H.~Elvang, D.~Z. Freedman, M.~Kiermaier, A.~Morales, {\em et al.},
  ``{E7(7) constraints on counterterms in N=8 supergravity},''
  \href{http://dx.doi.org/10.1016/j.physletb.2010.09.069}{{\em Phys.Lett.} {\bf
  B694} (2010)  265--271},
\href{http://arxiv.org/abs/1009.1643}{{\tt arXiv:1009.1643 [hep-th]}}.

\bibitem{Bossard:2011tq}
G.~Bossard, P.~Howe, K.~Stelle, and P.~Vanhove, ``{The vanishing volume of D=4
  superspace},'' \href{http://dx.doi.org/10.1088/0264-9381/28/21/215005}{{\em
  Class.Quant.Grav.} {\bf 28} (2011)  215005},
\href{http://arxiv.org/abs/1105.6087}{{\tt arXiv:1105.6087 [hep-th]}}.

\bibitem{Bossard:2012xs}
G.~Bossard, P.~S. Howe, and K.~S. Stelle, ``{Anomalies and divergences in N=4
  supergravity},'' \href{http://dx.doi.org/10.1016/j.physletb.2013.01.021}{{\em
  Phys. Lett.} {\bf B719} (2013)  424--429},
\href{http://arxiv.org/abs/1212.0841}{{\tt arXiv:1212.0841 [hep-th]}}.

\bibitem{Monteiro:2014cda}
R.~Monteiro, D.~O'Connell, and C.~D. White, ``{Black holes and the double
  copy},'' \href{http://dx.doi.org/10.1007/JHEP12(2014)056}{{\em JHEP} {\bf
  1412} (2014)  056},
\href{http://arxiv.org/abs/1410.0239}{{\tt arXiv:1410.0239 [hep-th]}}.

\bibitem{Luna:2015paa}
A.~Luna, R.~Monteiro, D.~O'Connell, and C.~D. White, ``{The classical double
  copy for Taub--NUT spacetime},''
  \href{http://dx.doi.org/10.1016/j.physletb.2015.09.021}{{\em Phys. Lett.}
  {\bf B750} (2015)  272--277},
\href{http://arxiv.org/abs/1507.01869}{{\tt arXiv:1507.01869 [hep-th]}}.

\bibitem{Luna:2016due}
A.~Luna, R.~Monteiro, I.~Nicholson, D.~O'Connell, and C.~D. White, ``{The
  double copy: Bremsstrahlung and accelerating black holes},''
  \href{http://dx.doi.org/10.1007/JHEP06(2016)023}{{\em JHEP} {\bf 06} (2016)
  023},
\href{http://arxiv.org/abs/1603.05737}{{\tt arXiv:1603.05737 [hep-th]}}.

\bibitem{Siegel:1988qu}
W.~Siegel, ``{SUPERSTRINGS GIVE OLD MINIMAL SUPERGRAVITY},''
\href{http://dx.doi.org/10.1016/0370-2693(88)90806-4}{{\em Phys.Lett.} {\bf
  B211} (1988)  55}.

\bibitem{Siegel:1995px}
W.~Siegel, ``{Curved extended superspace from Yang-Mills theory a la
  strings},'' \href{http://dx.doi.org/10.1103/PhysRevD.53.3324}{{\em Phys.Rev.}
  {\bf D53} (1996)  3324--3336},
\href{http://arxiv.org/abs/hep-th/9510150}{{\tt arXiv:hep-th/9510150
  [hep-th]}}.

\bibitem{Anastasiou:2014qba}
A.~Anastasiou, L.~Borsten, M.~J. Duff, L.~J. Hughes, and S.~Nagy, ``{Yang-Mills
  origin of gravitational symmetries},''
  \href{http://dx.doi.org/10.1103/PhysRevLett.113.231606}{{\em Phys.Rev.Lett.}
  {\bf 113} (2014) no.~23, 231606},
\href{http://arxiv.org/abs/1408.4434}{{\tt arXiv:1408.4434 [hep-th]}}.

\bibitem{Hodges:2011wm}
A.~Hodges, ``{New expressions for gravitational scattering amplitudes},''
  \href{http://dx.doi.org/10.1007/JHEP07(2013)075}{{\em Journal of High Energy
  Physics} {\bf 1307} (2013)  },
\href{http://arxiv.org/abs/1108.2227}{{\tt arXiv:1108.2227 [hep-th]}}.

\bibitem{Cachazo:2013iea}
F.~Cachazo, S.~He, and E.~Y. Yuan, ``{Scattering of Massless Particles:
  Scalars, Gluons and Gravitons},''
  \href{http://dx.doi.org/10.1007/JHEP07(2014)033}{{\em JHEP} {\bf 1407} (2014)
   033},
\href{http://arxiv.org/abs/1309.0885}{{\tt arXiv:1309.0885 [hep-th]}}.

\bibitem{Dolan:2013isa}
L.~Dolan and P.~Goddard, ``{Proof of the Formula of Cachazo, He and Yuan for
  Yang-Mills Tree Amplitudes in Arbitrary Dimension},''
  \href{http://dx.doi.org/10.1007/JHEP05(2014)010}{{\em JHEP} {\bf 1405} (2014)
   010},
\href{http://arxiv.org/abs/1311.5200}{{\tt arXiv:1311.5200 [hep-th]}}.

\bibitem{Sohnius:1981tp}
M.~F. Sohnius and P.~C. West, ``{An Alternative Minimal Off-Shell Version of
  N=1 Supergravity},''
\href{http://dx.doi.org/10.1016/0370-2693(81)90778-4}{{\em Phys.Lett.} {\bf
  B105} (1981)  353}.

\bibitem{Cecotti:1987qe}
S.~Cecotti, S.~Ferrara, M.~Porrati, and S.~Sabharwal, ``{NEW MINIMAL HIGHER
  DERIVATIVE SUPERGRAVITY COUPLED TO MATTER},''
\href{http://dx.doi.org/10.1016/0550-3213(88)90175-7}{{\em Nucl.Phys.} {\bf
  B306} (1988)  160}.

\bibitem{Ferrara:1988qxa}
S.~Ferrara and S.~Sabharwal, ``{Structure of New Minimal Supergravity},''
\href{http://dx.doi.org/10.1016/0003-4916(89)90167-X}{{\em Annals Phys.} {\bf
  189} (1989)  318--351}.

\bibitem{Stelle:1978ye}
K.~Stelle and P.~C. West, ``{Minimal Auxiliary Fields for Supergravity},''
\href{http://dx.doi.org/10.1016/0370-2693(78)90669-X}{{\em Phys.Lett.} {\bf
  B74} (1978)  330}.

\bibitem{Ferrara:1978em}
S.~Ferrara and P.~van Nieuwenhuizen, ``{The Auxiliary Fields of
  Supergravity},''
\href{http://dx.doi.org/10.1016/0370-2693(78)90670-6}{{\em Phys.Lett.} {\bf
  B74} (1978)  333}.

\bibitem{Hull:1994ys}
C.~M. Hull and P.~K. Townsend, ``{Unity of superstring dualities},''
  \href{http://dx.doi.org/10.1016/0550-3213(94)00559-W}{{\em Nucl. Phys.} {\bf
  B438} (1995)  109--137},
\href{http://arxiv.org/abs/hep-th/9410167}{{\tt arXiv:hep-th/9410167}}.

\bibitem{Cremmer:1979up}
E.~Cremmer and B.~Julia, ``{The $SO(8)$ supergravity},''
\href{http://dx.doi.org/10.1016/0550-3213(79)90331-6}{{\em Nucl. Phys.} {\bf
  B159} (1979)  141}.

\bibitem{Anastasiou:2015vba}
A.~Anastasiou, L.~Borsten, L.~J. Hughes, and S.~Nagy, ``{Global symmetries of
  Yang-Mills squared in various dimensions},'' {\em JHEP} {\bf 148} (2016)
  1601,
\href{http://arxiv.org/abs/1502.05359}{{\tt arXiv:1502.05359 [hep-th]}}.

\bibitem{Bianchi:2008pu}
M.~Bianchi, H.~Elvang, and D.~Z. Freedman, ``{Generating Tree Amplitudes in N=4
  SYM and N = 8 SG},''
  \href{http://dx.doi.org/10.1088/1126-6708/2008/09/063}{{\em JHEP} {\bf 0809}
  (2008)  063},
\href{http://arxiv.org/abs/0805.0757}{{\tt arXiv:0805.0757 [hep-th]}}.

\bibitem{Chiodaroli:2011pp}
M.~Chiodaroli, M.~Gunaydin, and R.~Roiban, ``{Superconformal symmetry and
  maximal supergravity in various dimensions},''
  \href{http://dx.doi.org/10.1007/JHEP03(2012)093}{{\em JHEP} {\bf 1203} (2012)
   093},
\href{http://arxiv.org/abs/1108.3085}{{\tt arXiv:1108.3085 [hep-th]}}.

\bibitem{Carrasco:2012ca}
J.~J.~M. Carrasco, M.~Chiodaroli, M.~G{\"u}naydin, and R.~Roiban, ``{One-loop
  four-point amplitudes in pure and matter-coupled N = 4 supergravity},''
  \href{http://dx.doi.org/10.1007/JHEP03(2013)056}{{\em JHEP} {\bf 1303} (2013)
   056},
\href{http://arxiv.org/abs/1212.1146}{{\tt arXiv:1212.1146 [hep-th]}}.

\bibitem{Chiodaroli:2014xia}
M.~Chiodaroli, M.~G{\"u}naydin, H.~Johansson, and R.~Roiban, ``{Scattering
  amplitudes in $ \mathcal{N}=2 $ Maxwell-Einstein and Yang-Mills/Einstein
  supergravity},'' \href{http://dx.doi.org/10.1007/JHEP01(2015)081}{{\em JHEP}
  {\bf 01} (2015)  081},
\href{http://arxiv.org/abs/1408.0764}{{\tt arXiv:1408.0764 [hep-th]}}.

\bibitem{Julia:1980gr}
B.~Julia, ``{Group disintegrations},'' in {\em Superspace and Supergravity},
  S.~Hawking and M.~Rocek, eds., vol.~C8006162 of {\em Nuffield Gravity
  Workshop}, pp.~331--350.
\newblock Cambridge University Press,
1980.
\newblock

\bibitem{Nicolai:1987kz}
H.~Nicolai, ``{The Integrability of $N=16$ Supergravity},''
\href{http://dx.doi.org/10.1016/0370-2693(87)91072-0}{{\em Phys.Lett.} {\bf
  B194} (1987)  402}.

\bibitem{West:2001as}
P.~C. West, ``{E(11) and M theory},''
  \href{http://dx.doi.org/10.1088/0264-9381/18/21/305}{{\em Class.Quant.Grav.}
  {\bf 18} (2001)  4443--4460},
\href{http://arxiv.org/abs/hep-th/0104081}{{\tt arXiv:hep-th/0104081
  [hep-th]}}.

\bibitem{Damour:2002cu}
T.~Damour, M.~Henneaux, and H.~Nicolai, ``{E(10) and a 'small tension
  expansion' of M theory},''
  \href{http://dx.doi.org/10.1103/PhysRevLett.89.221601}{{\em Phys.Rev.Lett.}
  {\bf 89} (2002)  221601},
\href{http://arxiv.org/abs/hep-th/0207267}{{\tt arXiv:hep-th/0207267
  [hep-th]}}.

\bibitem{Borsten:2013bp}
L.~Borsten, M.~J. Duff, L.~J. Hughes, and S.~Nagy, ``{A magic square from
  Yang-Mills squared},''
  \href{http://dx.doi.org/10.1103/PhysRevLett.112.131601}{{\em Phys.Rev.Lett.}
  {\bf 112} (2014)  131601},
\href{http://arxiv.org/abs/1301.4176}{{\tt arXiv:1301.4176 [hep-th]}}.

\bibitem{Freudenthal:1954}
H.~Freudenthal, ``{Beziehungen der $E_7$ und $E_8$ zur oktavenebene I-II},''
  {\em Nederl. Akad. Wetensch. Proc. Ser.} {\bf 57} (1954)  218--230.

\bibitem{Tits:1955}
J.~Tits, ``{Interpr\'{e}tation g\'{e}om\'{e}triques de groupes de Lie simples
  compacts de la classe $E$},'' {\em M\'{e}m. Acad. Roy. Belg. Sci} {\bf 29}
  (1955)  3.

\bibitem{Rosenfeld:1956}
B.~A. Rosenfeld, ``{Geometrical interpretation of the compact simple Lie groups
  of the class $E$ (Russian)},'' {\em Dokl. Akad. Nauk. SSSR} {\bf 106} (1956)
  600--603.

\bibitem{Cacciatori:2012cb}
S.~L. Cacciatori, B.~L. Cerchiai, and A.~Marrani, ``{Squaring the Magic},''
\href{http://arxiv.org/abs/1208.6153}{{\tt arXiv:1208.6153 [math-ph]}}.

\bibitem{Gunaydin:1983rk}
M.~G{\"u}naydin, G.~Sierra, and P.~K. Townsend, ``{Exceptional supergravity
  theories and the magic square},''
\href{http://dx.doi.org/10.1016/0370-2693(83)90108-9}{{\em Phys. Lett.} {\bf
  B133} (1983)  72}.

\bibitem{Gunaydin:1983bi}
M.~G{\"u}naydin, G.~Sierra, and P.~K. Townsend, ``{The geometry of $N=2$
  Maxwell-Einstein supergravity and Jordan algebras},''
\href{http://dx.doi.org/10.1016/0550-3213(84)90142-1}{{\em Nucl. Phys.} {\bf
  B242} (1984)  244}.

\bibitem{Gunaydin:1984ak}
M.~G{\"u}naydin, G.~Sierra, and P.~K. Townsend, ``{Gauging the $d = 5$
  Maxwell-Einstein supergravity theories: More on Jordan algebras},''
\href{http://dx.doi.org/10.1016/0550-3213(85)90547-4}{{\em Nucl. Phys.} {\bf
  B253} (1985)  573}.

\bibitem{Barton:2003}
C.~H. Barton and A.~Sudbery, ``{Magic squares and matrix models of Lie
  algebras},'' \href{http://dx.doi.org/10.1016/S0001-8708(03)00015-X}{{\em Adv.
  in Math.} {\bf 180} (2003) no.~2, 596--647},
  \href{http://arxiv.org/abs/math/0203010}{{\tt arXiv:math/0203010}}.

\bibitem{Baez:2001dm}
J.~C. Baez, ``{The Octonions},''
  \href{http://dx.doi.org/10.1090/S0273-0979-01-00934-X}{{\em Bull. Am. Math.
  Soc.} {\bf 39} (2002)  145--205},
\href{http://arxiv.org/abs/math/0105155}{{\tt arXiv:math/0105155 [math-ra]}}.

\bibitem{Hurwitz:1898}
A.~Hurwitz, ``{Uber die komposition der quadratishen formen von beliebig vielen
  variabeln},'' {\em Nachr. Ges. Wiss. Gottingen} (1898)  309--316.

\bibitem{Moufang:1933}
R.~Moufang, ``Alternativk\"{o}rper und der satz vom vollsta ̈ndigen
  vierseit,'' {\em Abh. Math. Sem. Hamburg} {\bf 9} (1933)  207--222.

\bibitem{Jordan:1949}
P.~Jordan, ``\"{U}ber eine nicht-desarguessche ebene projektive geometrie,''
  {\em Abh. Math. Sem. Hamburg} {\bf 16} (1949)  74--76.

\bibitem{Chevally:1950}
C.~Chevalley and R.~D. Schafer, ``The exceptional simple lie algebras $f_4$ and
  $e_6$,'' {\em Proc. Nat. Acad. Sci. USA} (1950)  137--141.

\bibitem{Anastasiou:2013hba}
A.~Anastasiou, L.~Borsten, M.~J. Duff, L.~J. Hughes, and S.~Nagy, ``{A magic
  pyramid of supergravities},''
  \href{http://dx.doi.org/10.1007/JHEP04(2014)178}{{\em JHEP} {\bf 1404} (2014)
   178},
\href{http://arxiv.org/abs/1312.6523}{{\tt arXiv:1312.6523 [hep-th]}}.

\bibitem{Gunaydin:1975mp}
M.~Gunaydin, ``{Exceptional Realizations of Lorentz Group: Supersymmetries and
  Leptons},''
\href{http://dx.doi.org/10.1007/BF02734524}{{\em Nuovo Cim.} {\bf A29} (1975)
  467}.

\bibitem{Gunaydin:1976vq}
M.~Gunaydin, ``{Octonionic Hilbert Spaces, the Poincare Group and SU(3)},''
\href{http://dx.doi.org/10.1063/1.522811}{{\em J.Math.Phys.} {\bf 17} (1976)
  1875}.

\bibitem{Gursey:1978et}
F.~Gursey, ``{Octonionic Structures in Particle Physics},''
{\em Lect.Notes Phys.} {\bf 94} (1979)  508--521.

\bibitem{Gunaydin:1979df}
M.~Gunaydin, ``{Quadratic Jordan Formulation Of Quantum Mechanics And
  Construction Of Lie (Super)Algebras From Jordan (Super)Algebras},'' in {\em
  International colloquium on group theoretical methods in physics}, vol.~10,
  p.~18.
\newblock Israel Grp.Th.Meth.,
1979.
\newblock

\bibitem{Kugo:1982bn}
T.~Kugo and P.~K. Townsend, ``{Supersymmetry and the division algebras},''
\href{http://dx.doi.org/10.1016/0550-3213(83)90584-9}{{\em Nucl. Phys.} {\bf
  B221} (1983)  357}.

\bibitem{Sudbery:1984}
A.~Sudbery, ``{Division algebras, (pseudo)orthogonal groups, and spinors},''
  \href{http://dx.doi.org/10.1088/0305-4470/17/5/018}{{\em J. Phys.} {\bf A17}
  (1984) no.~5, 939--955}.

\bibitem{Gursey:1987mv}
F.~Gursey, ``{Superpoincare Groups and Division Algebras},''
\href{http://dx.doi.org/10.1142/S0217732387001221}{{\em Mod.Phys.Lett.} {\bf
  A2} (1987)  967}.

\bibitem{Green:1987sp}
M.~B. Green, J.~H. Schwarz, and E.~Witten, {\em {Superstring Theory vol. 1:
  Introduction}}.
\newblock Cambridge Monographs on Mathematical Physics. Cambridge University
  Press, Cambridge, UK, 1987.
\newblock 469 p.

\bibitem{Evans:1987tm}
J.~M. Evans, ``{Supersymmetric Yang-Mills theories and division algebras},''
\href{http://dx.doi.org/10.1016/0550-3213(88)90305-7}{{\em Nucl. Phys.} {\bf
  B298} (1988)  92}.

\bibitem{Duff:1987qa}
M.~J. Duff, ``{Supermembranes: The First Fifteen Weeks},''
\href{http://dx.doi.org/10.1088/0264-9381/5/1/023}{{\em Class.Quant.Grav.} {\bf
  5} (1988)  189}.

\bibitem{Blencowe:1988sk}
M.~Blencowe and M.~J. Duff, ``{Supermembranes and the Signature of
  Space-time},''
\href{http://dx.doi.org/10.1016/0550-3213(88)90155-1}{{\em Nucl.Phys.} {\bf
  B310} (1988)  387}.

\bibitem{Gunaydin:1992zh}
M.~Gunaydin, ``{Generalized conformal and superconformal group actions and
  Jordan algebras},'' \href{http://dx.doi.org/10.1142/S0217732393001124}{{\em
  Mod.Phys.Lett.} {\bf A8} (1993)  1407--1416},
\href{http://arxiv.org/abs/hep-th/9301050}{{\tt arXiv:hep-th/9301050
  [hep-th]}}.

\bibitem{Berkovits:1993hx}
N.~Berkovits, ``{A Ten-dimensional superYang-Mills action with off-shell
  supersymmetry},'' \href{http://dx.doi.org/10.1016/0370-2693(93)91791-K}{{\em
  Phys.Lett.} {\bf B318} (1993)  104--106},
\href{http://arxiv.org/abs/hep-th/9308128}{{\tt arXiv:hep-th/9308128
  [hep-th]}}.

\bibitem{Manogue:1993ja}
C.~A. Manogue and J.~Schray, ``{Finite Lorentz transformations, automorphisms,
  and division algebras},'' \href{http://dx.doi.org/10.1063/1.530056}{{\em J.
  Math. Phys.} {\bf 34} (1993)  3746--3767},
\href{http://arxiv.org/abs/hep-th/9302044}{{\tt arXiv:hep-th/9302044}}.

\bibitem{Evans:1994cn}
J.~M. Evans, ``{Auxiliary fields for superYang-Mills from division algebras},''
  {\em Lect.Notes Phys.} {\bf 447} (1995)  218--223,
\href{http://arxiv.org/abs/hep-th/9410239}{{\tt arXiv:hep-th/9410239
  [hep-th]}}.

\bibitem{Schray:1994ur}
J.~Schray and C.~A. Manogue, ``{Octonionic representations of Clifford algebras
  and triality},'' \href{http://dx.doi.org/10.1007/BF02058887}{{\em Found.
  Phys.} {\bf 26} (1996) no.~1, 17--70},
\href{http://arxiv.org/abs/hep-th/9407179}{{\tt arXiv:hep-th/9407179}}.

\bibitem{gursey1996role}
F.~G{\"u}rsey and C.-H. Tze,
  \href{http://dx.doi.org/10.1142/9789812819857}{{\em On the role of division,
  Jordan and related algebras in particle physics}}.
\newblock World Scientific, London, 1996.
\newblock \url{http://www.worldscientific.com/doi/abs/10.1142/9789812819857}.

\bibitem{Manogue:1998rv}
C.~A. Manogue and T.~Dray, ``{Dimensional reduction},''
  \href{http://dx.doi.org/10.1142/S0217732399000134}{{\em Mod.Phys.Lett.} {\bf
  A14} (1999)  99--104},
\href{http://arxiv.org/abs/hep-th/9807044}{{\tt arXiv:hep-th/9807044
  [hep-th]}}.

\bibitem{Gunaydin:2000xr}
M.~G{\"u}naydin, K.~Koepsell, and H.~Nicolai, ``{Conformal and quasiconformal
  realizations of exceptional Lie groups},''
  \href{http://dx.doi.org/10.1007/PL00005574}{{\em Commun. Math. Phys.} {\bf
  221} (2001)  57--76},
\href{http://arxiv.org/abs/hep-th/0008063}{{\tt arXiv:hep-th/0008063}}.

\bibitem{Toppan:2003yx}
F.~Toppan, ``{On the octonionic M-superalgebra},'' in {\em Sao Paulo 2002,
  Integrable theories, solitons and duality}.
\newblock 2002.
\newblock
\href{http://arxiv.org/abs/hep-th/0301163}{{\tt arXiv:hep-th/0301163}}.
\newblock

\bibitem{Gunaydin:2005zz}
M.~G{\"u}naydin and O.~Pavlyk, ``{Generalized spacetimes defined by cubic forms
  and the minimal unitary realizations of their quasiconformal groups},''
  \href{http://dx.doi.org/10.1088/1126-6708/2005/08/101}{{\em JHEP} {\bf 08}
  (2005)  101},
\href{http://arxiv.org/abs/hep-th/0506010}{{\tt arXiv:hep-th/0506010}}.

\bibitem{Borsten:2008wd}
L.~Borsten, D.~Dahanayake, M.~J. Duff, H.~Ebrahim, and W.~Rubens, ``{Black
  Holes, Qubits and Octonions},''
  \href{http://dx.doi.org/10.1016/j.physrep.2008.11.002}{{\em Phys. Rep.} {\bf
  471} (2009) no.~3--4, 113--219},
\href{http://arxiv.org/abs/0809.4685}{{\tt arXiv:0809.4685 [hep-th]}}.

\bibitem{Baez:2009xt}
J.~C. Baez and J.~Huerta, ``Division algebras and supersymmetry i,'' in {\em
  Superstrings, Geometry, Topology, and C*-Algebras, eds. R. Doran, G. Friedman
  and J. Rosenberg, Proc. Symp. Pure Math}, vol.~81, pp.~65--80.
\newblock 2009.
\newblock
\href{http://arxiv.org/abs/0909.0551}{{\tt arXiv:0909.0551 [hep-th]}}.
\newblock

\bibitem{Baez:2010ye}
J.~C. Baez and J.~Huerta, ``{Division Algebras and Supersymmetry II},''
  \href{http://dx.doi.org/10.4310/ATMP.2011.v15.n5.a4}{{\em Adv. Theor. Math.
  Phys.} {\bf 15} (2011) no.~5, 1373--1410},
\href{http://arxiv.org/abs/1003.3436}{{\tt arXiv:1003.3436 [hep-th]}}.

\bibitem{Huerta:2011ic}
J.~Huerta, ``{Division Algebras, Supersymmetry and Higher Gauge Theory},''
\href{http://arxiv.org/abs/1106.3385}{{\tt arXiv:1106.3385 [math-ph]}}.

\bibitem{Huerta:2011aa}
J.~Huerta, ``{Division Algebras and Supersymmetry III},''
  \href{http://dx.doi.org/10.4310/ATMP.2012.v16.n5.a4}{{\em
  Adv.Theor.Math.Phys.} {\bf 16} (2012)  1485--1589},
\href{http://arxiv.org/abs/1109.3574}{{\tt arXiv:1109.3574 [hep-th]}}.

\bibitem{Anastasiou:2014zfa}
A.~Anastasiou, L.~Borsten, M.~J. Duff, L.~J. Hughes, and S.~Nagy, ``{An
  octonionic formulation of the M-theory algebra},''
  \href{http://dx.doi.org/10.1007/JHEP11(2014)022}{{\em JHEP} {\bf 1411} (2014)
   022},
\href{http://arxiv.org/abs/1402.4649}{{\tt arXiv:1402.4649 [hep-th]}}.

\bibitem{Huerta:2014loa}
J.~Huerta, ``{Division Algebras and Supersymmetry IV},''
\href{http://arxiv.org/abs/1409.4361}{{\tt arXiv:1409.4361 [hep-th]}}.

\bibitem{Marrani:2014qia}
A.~Marrani and P.~Truini, ``{Exceptional Lie Algebras, SU(3) and Jordan Pairs
  Part 2: Zorn-type Representations},''
  \href{http://dx.doi.org/10.1088/1751-8113/47/26/265202}{{\em J.Phys.} {\bf
  A47} (2014)  265202},
\href{http://arxiv.org/abs/1403.5120}{{\tt arXiv:1403.5120 [math-ph]}}.

\bibitem{Chung:1987in}
K.~Chung and A.~Sudbery, ``{Octonions and the Lorentz and Conformal Groups of
  Ten-dimensional Space-time},''
\href{http://dx.doi.org/10.1016/0370-2693(87)91489-4}{{\em Phys.Lett.} {\bf
  B198} (1987)  161}.

\bibitem{Fairlie:1987td}
D.~B. Fairlie and C.~A. Manogue, ``{A Parametrization of the Covariant
  Superstring},''
\href{http://dx.doi.org/10.1103/PhysRevD.36.475}{{\em Phys.Rev.} {\bf D36}
  (1987)  475}.

\bibitem{Manogue:1989ey}
C.~A. Manogue and A.~Sudbery, ``{General Solutions of Covariant Superstring
  Equations of Motion},''
\href{http://dx.doi.org/10.1103/PhysRevD.40.4073}{{\em Phys.Rev.} {\bf D40}
  (1989)  4073}.

\bibitem{Schray:1994fc}
J.~Schray, ``{The General classical solution of the superparticle},''
  \href{http://dx.doi.org/10.1088/0264-9381/13/1/004}{{\em Class.Quant.Grav.}
  {\bf 13} (1996)  27--38},
\href{http://arxiv.org/abs/hep-th/9407045}{{\tt arXiv:hep-th/9407045
  [hep-th]}}.

\bibitem{dray2000octonionic}
T.~Dray, J.~Janesky, and C.~A. Manogue, ``Octonionic hermitian matrices with
  non-real eigenvalues,'' {\em Advances in Applied Clifford Algebras} {\bf 10}
  (2000) no.~2, 193--216.

\bibitem{Anastasiou:2013cya}
A.~Anastasiou, L.~Borsten, M.~J. Duff, L.~J. Hughes, and S.~Nagy, ``{Super
  Yang-Mills, division algebras and triality},''
  \href{http://dx.doi.org/10.1007/JHEP08(2014)080}{{\em JHEP} {\bf 1408} (2014)
   080},
\href{http://arxiv.org/abs/1309.0546}{{\tt arXiv:1309.0546 [hep-th]}}.

\bibitem{Hull:2000zn}
C.~Hull, ``{Strongly coupled gravity and duality},''
  \href{http://dx.doi.org/10.1016/S0550-3213(00)00323-0}{{\em Nucl.Phys.} {\bf
  B583} (2000)  237--259},
\href{http://arxiv.org/abs/hep-th/0004195}{{\tt arXiv:hep-th/0004195
  [hep-th]}}.

\bibitem{Hull:2000rr}
C.~Hull, ``{Symmetries and compactifications of (4,0) conformal gravity},''
  \href{http://dx.doi.org/10.1088/1126-6708/2000/12/007}{{\em JHEP} {\bf 0012}
  (2000)  007},
\href{http://arxiv.org/abs/hep-th/0011215}{{\tt arXiv:hep-th/0011215
  [hep-th]}}.

\bibitem{Hull:2000ih}
C.~Hull, ``{Conformal nongemetric gravity in six-dimensions and M theory above
  the Planck energy},''
  \href{http://dx.doi.org/10.1088/0264-9381/18/16/313}{{\em Class.Quant.Grav.}
  {\bf 18} (2001)  3233--3240},
\href{http://arxiv.org/abs/hep-th/0011171}{{\tt arXiv:hep-th/0011171
  [hep-th]}}.

\bibitem{Huang:2010rn}
Y.-t. Huang and A.~E. Lipstein, ``{Amplitudes of 3D and 6D Maximal
  Superconformal Theories in Supertwistor Space},''
  \href{http://dx.doi.org/10.1007/JHEP10(2010)007}{{\em JHEP} {\bf 1010} (2010)
   007},
\href{http://arxiv.org/abs/1004.4735}{{\tt arXiv:1004.4735 [hep-th]}}.

\bibitem{Czech:2011dk}
B.~Czech, Y.-t. Huang, and M.~Rozali, ``{Chiral three-point interactions in 5
  and 6 dimensions},'' \href{http://dx.doi.org/10.1007/JHEP10(2012)143}{{\em
  JHEP} {\bf 1210} (2012)  143},
\href{http://arxiv.org/abs/1110.2791}{{\tt arXiv:1110.2791 [hep-th]}}.

\bibitem{Bargheer:2012gv}
T.~Bargheer, S.~He, and T.~McLoughlin, ``{New Relations for Three-Dimensional
  Supersymmetric Scattering Amplitudes},''
  \href{http://dx.doi.org/10.1103/PhysRevLett.108.231601}{{\em Phys.Rev.Lett.}
  {\bf 108} (2012)  231601},
\href{http://arxiv.org/abs/1203.0562}{{\tt arXiv:1203.0562 [hep-th]}}.

\bibitem{Huang:2012wr}
Y.-t. Huang and H.~Johansson, ``{Equivalent D=3 Supergravity Amplitudes from
  Double Copies of Three-Algebra and Two-Algebra Gauge Theories},''
  \href{http://dx.doi.org/10.1103/PhysRevLett.110.171601}{{\em Phys.Rev.Lett.}
  {\bf 110} (2013)  171601},
\href{http://arxiv.org/abs/1210.2255}{{\tt arXiv:1210.2255 [hep-th]}}.

\bibitem{Bagger:2006sk}
J.~Bagger and N.~Lambert, ``{Modeling Multiple M2's},''
  \href{http://dx.doi.org/10.1103/PhysRevD.75.045020}{{\em Phys.Rev.} {\bf D75}
  (2007)  045020},
\href{http://arxiv.org/abs/hep-th/0611108}{{\tt arXiv:hep-th/0611108
  [hep-th]}}.

\bibitem{Gustavsson:2007vu}
A.~Gustavsson, ``{Algebraic structures on parallel M2-branes},''
  \href{http://dx.doi.org/10.1016/j.nuclphysb.2008.11.014}{{\em Nucl.Phys.}
  {\bf B811} (2009)  66--76},
\href{http://arxiv.org/abs/0709.1260}{{\tt arXiv:0709.1260 [hep-th]}}.

\bibitem{Bagger:2007jr}
J.~Bagger and N.~Lambert, ``{Gauge symmetry and supersymmetry of multiple
  M2-branes},'' \href{http://dx.doi.org/10.1103/PhysRevD.77.065008}{{\em
  Phys.Rev.} {\bf D77} (2008)  065008},
\href{http://arxiv.org/abs/0711.0955}{{\tt arXiv:0711.0955 [hep-th]}}.

\bibitem{Borsten:2015toappear}
L.~Borsten, ``On the $d=6, \mathcal{N}=(4, 0)$ theory.'' to appear, 2016.

\end{thebibliography}


\providecommand{\href}[2]{#2}\begingroup\raggedright\endgroup

\end{document}